\documentclass[showpacs,twocolumn,prx,superscriptaddress]{revtex4-1}

\usepackage[utf8]{inputenc} 

\usepackage{qcircuit}
\usepackage[dvips]{graphicx}
\usepackage{siunitx}
\usepackage{amsmath,amssymb,amsthm,mathrsfs,amsfonts,dsfont}
\usepackage{subfigure, epsfig}
\usepackage{braket}
\usepackage{bm}
\usepackage{enumerate}
\usepackage[dvipsnames]{xcolor}
\usepackage{color}
\usepackage{fancybox, graphicx}
\usepackage[skins]{tcolorbox}
\usepackage{multirow}
\usepackage{mathtools}

\usepackage[colorlinks]{hyperref}
\hypersetup{
	colorlinks=true,
	citecolor = {blue},
    linkcolor=blue,
    filecolor=magenta,      
    urlcolor=cyan,
}

\newtcolorbox[auto counter]{tbox}[2][]{%
	enhanced, float=hbt, drop fuzzy shadow southeast,
	colback=white!5!white, colframe=white!30!black,
	width= .97\columnwidth,sharp corners,boxrule=0.8pt,
	title={#2}, #1
}

\newtcolorbox{codebox}{enhanced, width=.95\columnwidth, halign = flush left, drop fuzzy shadow southeast, boxrule=0.4pt, sharp corners, colframe=black, colback=white}
\begin{document}

\title{Learning from physics experiments, with quantum computers: \\
Applications in muon spectroscopy}

\date{\today}
\author{Sam McArdle}
\email{sam.mcardle.science@gmail.com}
\affiliation{Department of Materials, University of Oxford, Parks Road, Oxford OX1 3PH, United Kingdom}

\maketitle

\textbf{Computational physics is an important tool for analysing, verifying, and -- at times -- replacing physical experiments. Nevertheless, simulating quantum systems and analysing quantum data has so far resisted an efficient classical treatment in full generality. While programmable quantum systems have been developed to address this challenge~\cite{RevModPhys.86.153}, the resources required for classically intractable problems still lie beyond our reach~\cite{childs2017toward, babbush2018encoding}. In this work, we consider a new target for quantum simulation algorithms; analysing the data arising from physics experiments - specifically, muon spectroscopy experiments. These experiments can be used to probe the quantum interactions present in condensed matter systems~\cite{yaouanc2011MuonTextbook}. However, fully analysing their results can require classical computational resources scaling exponentially with the simulated system size, which can limit our understanding of the studied system. We show that this task may be a natural fit for the coming generations of quantum computers. We use classical emulations of our quantum algorithm on systems of up to 29 qubits to analyse real experimental data, and to estimate both the near-term and error corrected resources required for our proposal. We find that our algorithm exhibits good noise resilience, stemming from our desire to extract global parameters from a fitted curve, rather than targeting any individual data point. In some respects, our resource estimates go further than some prior work in quantum simulation, by estimating the resources required to solve a complete task, rather than just to run a given circuit. Taking the overhead of observable measurement and calculating multiple datapoints into account, we find that significant challenges still remain if our algorithm is to become practical for analysing muon spectroscopy data.}\\

At first glance, quantum computing appears to offer remarkable computational power; potentially yielding exponential speedups for simulating quantum systems, or for solving some problems in machine learning. In reality, however, the situation is more nuanced, and there are a number of challenges that quantum algorithms must overcome, in order to become practical for real problems of interest. At the heart of these challenges, is that we must first build a quantum computer that is both sufficiently large, and has low enough error rates to run the calculation (in an error corrected setting these essentially become the same requirement, as the error rate can be suppressed by adding additional qubits for increased error protection). How large and noiseless the computer must be is determined by the efficiency of the proposed quantum algorithms, as well as the efficiency of the classical algorithms they aim to surpass. 

The most promising use cases for quantum simulation and machine learning algorithms seem to require a number of features. Firstly, the problem should be hard to solve classically, and should not admit accurate classical approximations. Quantum algorithms for solving the electronic structure problem must overcome this challenge, as sophisticated classical methods can accurately approximate the low-lying energy levels of many small systems of interest, up to around 100 qubits~\cite{simonscollab2015hubbard,simons2020manybodycomp}. Secondly, the output of the algorithm should be resilient to noise, which will reduce the resources required when performing quantum error correction, or make the calculation amenable to noisy, near-term quantum hardware. In the case of quantum machine learning algorithms, the problem should also have an inherently quantum structure, to avoid relying on heuristic arguments for quantum advantage. The model should be efficient to train, which is aided by having a good prior for the model parameters and structure~\cite{Barren,cerezo2020cost}. Finally, there should be no data input/output limitations, which avoids the need for hypothetical data structures like quantum random access memory, or the ability to directly interface a quantum computer with other quantum systems which supply the input data state, such as the interactive quantum likelihood
evaluation method~\cite{wiebe2014hamiltonianlearning} or quantum convolutional neural networks~\cite{cong2019quantum}. \\

\begin{figure*}
\includegraphics[width=1.8\columnwidth]{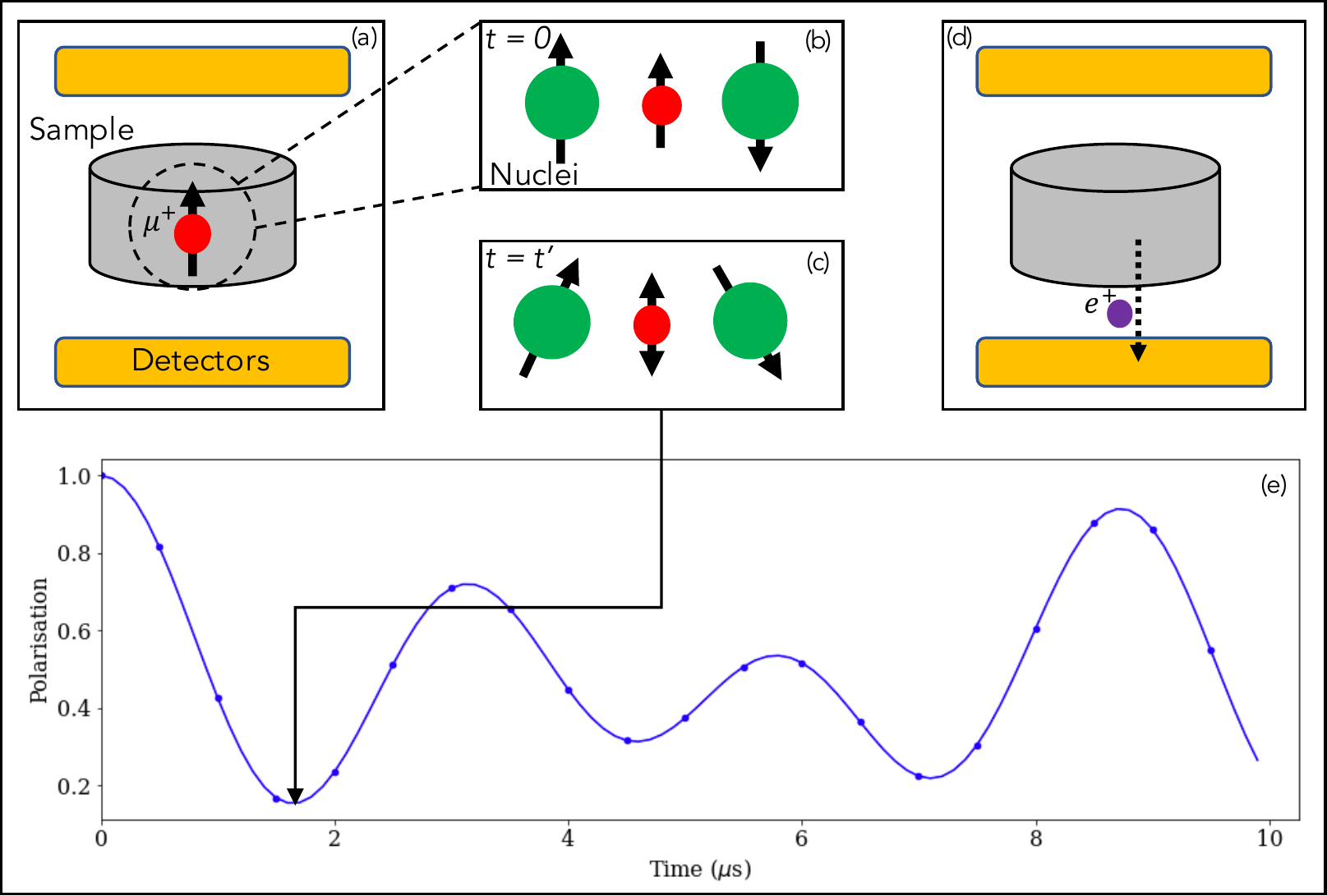}
 \caption{A schematic of a muon spectroscopy experiment, as described in the main text. Spin polarised muons are implanted into a sample of interest (a), whereupon they interact with nearby nuclear spins (b). This causes an oscillation of the muon spin polarisation (c). When the muon decays, the positron is preferentially emitted in the direction of the muon spin at the time that it decayed (d). By recording the normalised asymmetry between positron counts registered by the two detectors, we can reconstruct the time evolution of the muon polarisation (e).} \label{Fig:Exp_diagram}
\end{figure*}

In this work, we introduce a quantum algorithm that sits at the interface of quantum simulation and quantum machine learning, which we apply to analysing the spectra arising from muon spin rotation, resonance, and relaxation ($\mu^+$SR) experiments. We have sought to optimise our algorithm against the criteria listed above, in order to ensure its practicality for solving real problems of interest. Muon spectroscopy experiments have been used to analyse a wide range of physical systems and phenomena, such as superconductivity and magnetism. These experiments typically measure the time evolution of the spin polarisation of anti-muons that have been implanted into a sample of interest~\cite{yaouanc2011MuonTextbook}. A typical experimental setup is shown in Fig.~\ref{Fig:Exp_diagram} and discussed fully in Sec.~\ref{Sec:MuonSpectro}. As the muon spin interacts with other environmental spins in the system, it oscillates between eigenstates of the system Hamiltonian. Tracking these variations enables us to learn quantitatively about the interactions at the muon site. Although the inputs to the model are Hamiltonian parameters -- classical data -- and the outputs of the experiments are classical data points, accurate analysis can require a fully quantum treatment of the system dynamics. In some cases, this problem appears challenging to solve on classical computers, admitting few simplifications, and often requiring exact diagonalisation of the system Hamiltonian~\cite{Celio1984ExactMuPolar, wilkinson2020muonfluorinefull,lord2000StrongCoupledMuExact}.

In Sec.~\ref{Sec:QC_for_muons} we present an algorithm to solve this problem efficiently using a quantum computer. Our algorithm is relatively simple, and requires modest computational resources. In particular, the finite lifetime of the anti-muon (\SI{2.2}{\micro\second} on average)
and our desire to extract global system properties appears to set a generous limit on the precision required from our quantum calculations. In Sec.~\ref{Sec:Results} we carry out classical numerical emulations of our quantum algorithm, to investigate its scaling behaviour and noise robustness. We use these emulations to analyse muon spectroscopy data from a real experiment, finding good agreement with recent state-of-the-art classical analysis~\cite{wilkinson2020muonfluorinefull}. The calculations performed as part of our analysis, involving a Hilbert space dimension of $2^{29}$, are the largest to date in the muon literature. We use these numerical results as the basis for estimating both the near-term and error corrected resources required to run our algorithm for problem sizes of interest. At first glance, our algorithm appears to require fewer fault tolerant resources than solving challenging instances of the electronic structure problem. However, the large number of repetitions required by our algorithm (to estimate observables, calculate multiple datapoints, and repeat the calculation as part of an optimisation loop) may result in an impractically long runtime. We argue in Sec.~\ref{Sec:Conclude} that this is not necessarily a limitation of our algorithm, but a challenge facing many quantum algorithms.

Our approach is similar in nature to the proposals of Refs.~\cite{chiesa2019quantum, sels2019quantum}, which also consider using quantum computers to analyse the outputs of quantum experiments (inelastic neutron scattering on simplified models of magnetic molecules, and linear response nuclear magnetic resonance experiments, respectively). All three methods are related to quantum likelihood evaluation~\cite{wiebe2014hamiltonianlearning}, whereby the data from one untrusted quantum source (the experiment) is fitted to simulated data from a trusted quantum source (the quantum computer) in order to extract system parameters of interest.

\section{Muon spectroscopy}\label{Sec:MuonSpectro}
Muon spectroscopy, more commonly known as muon spin rotation, relaxation and resonance ($\mu^+$SR), emerged as an experimental technique in the 1970s~\cite{gurevich1972MuonCopperFirst}, following the theoretical proposal by \textcite{garwin1957MuonProposal}. The technique is closely related to other spin-based methods for probing magnetic interactions, such as nuclear magnetic resonance (NMR) and electron spin resonance (ESR). $\mu^+$SR experiments can provide quantitative information on a range of physical phenomena, including: magnetic ordering and phase transitions in materials, the diffusion of light interstitial defects in semiconductors, vortex formation in superconductors, and low dimensional magnetism. In this section, we provide an introduction to $\mu^+$SR. We direct the interested reader to the review article by \textcite{blundell1999muons} and the textbook by \textcite{yaouanc2011MuonTextbook} for more detailed discussions of the method.

As mentioned in the introduction, $\mu^+$SR uses spin-polarised beams of positive (anti)-muons (hereafter referred to as muons, as the ordinary muon is rarely used in muon spectroscopy experiments, and so will not be discussed in this work) to probe the interactions in a sample of interest. A typical $\mu^+$SR experiment proceeds as follows:
\begin{itemize}
    \item The $\mu^+$ is a positively charged, spin-$\frac{1}{2}$ particle, with a mass approximately $1/9^\mathrm{th}$ that of the proton. It can be produced at particle accelerators, from the decay of pions: $\pi^+ \rightarrow \mu^+ + \nu_\mathrm{\mu}$. Working in the rest frame of the pion, and conserving linear and angular momentum, we find that both the muon emission direction and spin polarisation (the spin projection along a specified axis) must be opposite to those of the neutrino. This decay proceeds via the weak interaction, and so violates parity. This causes the neutrino to have its spin polarisation orientated anti-parallel to its momentum - and thus, the same happens to the muon. By selecting muons arising from pions decaying at rest, spin-polarised muon beams are produced. In this work, we choose the initial spin-polarisation of the muon beam to be along the positive $Z$ axis. As a result, we can consider the muons to initially be in state $\ket{0}\bra{0}_\mu$, where $\ket{0}$ is the +1 eigenstate of the Pauli $Z$ matrix.
    
    \item The beam is directed into the sample of interest. The muons are brought to rest by electrostatic interactions, which do not depolarise the beam~\cite{blundell1999muons}.
    
    \item The muons interact with their local environment via spin-spin interactions. As the initial polarisation of the muon spin is unlikely to be an eigenstate of the muon-environment Hamiltonian, the muon spin polarisation will evolve according to the Schr\"odinger equation.
    
    \item The muon decays with a half-life of \SI{2.2}{\micro\second} into a positron, electron neutrino and muon antineutrino. Once again, as a consequence of the weak interaction, the positron is emitted preferentially in the direction of the muon spin polarisation. 
    
    \item The emitted positrons exit the sample, and are detected by detectors placed forwards and backwards along the direction of the muons' initial spin-polarisation. The detection of positrons gives an asymmetry function
    \begin{equation}\label{Eq:Asymmetry}
        A(t) = \frac{N_+ - \alpha N_-}{N_+ + \alpha N_-}
    \end{equation}
    which records the normalised difference between the forward and backwards detector counts ($N_{+/-}$, respectively), where $\alpha$ accounts for detection inefficiencies and imbalances. This value is converted to a polarisation function, using 
    \begin{equation}\label{Eq:Asym_to_pol}
        P(t) = \frac{A(t) - A_\mathrm{bg}}{A_0}
    \end{equation}
    where $A_\mathrm{bg}$ gives the background positron count, and $A_0 = A(0) - A_\mathrm{bg}$. The polarisation function is normalised to between $\pm 1$, and corresponds to the spin-polarisation of the muon beam at time $t$. Mathematically, the polarisation function is given by
    \begin{equation}\label{Eq:Pol}
    \begin{aligned}
        P(t) &= \mathrm{Tr}\big{(}Z_\mu \rho_s(t)\big{)} \\ &= \mathrm{Tr}\big{(}Z_\mu e^{-iH_st} \big{[} \ket{0}\bra{0}_\mu \otimes \rho_e(0)\big{]} e^{iH_st}\big{)},
    \end{aligned}
    \end{equation}
    where $Z_\mu$ is the Pauli $Z$ matrix acting on the muon, $H_s$ is the system Hamiltonian, $\rho_s(t)$ is the state of the muon-environment system at time $t$, and $\rho_e(0)$ is the initial state of the environment.
    
    \item Measuring the time evolution of the polarisation function enables us to infer the interactions felt by the muon at its rest site (given a model for the initial state of the environment), and so learn quantitatively about the sample.
\end{itemize}

An understanding of $\mu^+$SR experiments can be gleaned from the following semi-classical example. A spin-polarised muon beam is directed into a sample of interest, where it interacts with a transverse magnetic field $\vec{B}$. The muon spin polarisation will precess at a frequency $\omega = B\gamma_\mu$, where $\gamma_\mu = 2\pi \times 135.5$~MHz$\cdot$T$^{-1}$ is the gyromagnetic ratio of the muon. After converting the positron counts into the polarisation function, we will observe that the polarisation function is given by $P(t) = \mathrm{cos}(B \gamma_\mu t)$. This enables us to infer the strength of the magnetic field at the muon site. This example is similar in spirit to many of the early $\mu^+$SR experiments performed. Due to their large magnetic moment, muons can be used as sensitive probes of both local magnetic fields and spin-spin interactions. \\

There are two types of muon beams: continuous wave, and pulsed. Muons in a continous beam are implanted one by one; a timer is started when a muon enters the sample, and stopped when a positron is detected. If the previous muon has not decayed before the next muon arrives, then these events are discarded. In order to keep the count rate high, the polarisation function is only recorded to around \SI{10}{\micro\second}. Continuous beams have a time resolution of around \SI{1}{\nano\second}. In contrast, pulsed beams implant hundreds of muons into the sample in a single pulse. The arrival time between pulses is large compared to the muon lifetime, which enables polarisation functions to be measured to around \SI{30}{\micro\second}. The resolution of pulsed beams is limited to around \SI{0.1}{\micro\second} by the finite width of the pulses. As a result, continuous and pulsed beams are better suited for studying fast and slow dynamics, respectively. 

Experiments can be carried out at high pressures ($\sim 1.5$~GPa), low temperatures (less than 1~K), with high strength transverse or longitudinal magnetic fields ($\sim 8$~T), or with the application of time-dependent radio-frequency pulses~\cite{yaouanc2011MuonTextbook}. Each of these techniques enables the investigation of certain phenomena more closely. For example, if we are investigating relaxation of the muon spin polarisation due to either T$_1$ relaxation (arising from magnetic field fluctuations in time, which cause the muon to exchange energy with the environment) or T$_2$ relaxation (caused by a spatially varying static field distribution, which causes the muons to precess at different frequencies, dephasing the beam) then a strong longitudinal magnetic field can be used. This `locks' the muon polarisation along its initial direction, which then makes it easier to measure differences in the polarisation function arising from relaxation processes. \\

As described above, $\mu^+$SR experiments have been used to investigate a range of physical phenomena. For example, we can measure the temperature dependence of oscillations in the polarisation function (or the lack thereof) in order to investigate phase transitions in magnetic materials, such as low dimensional spin chains~\cite{lancaster2019mu_spin_chain}. Other experiments have measured the polarisation function of muons in semiconductors, at a range of temperatures, in order to investigate the diffusion of muons within the sample. The muon acts as a light proton (often capturing an electron in semiconductors to form `muonium'), so these experiments can be used to examine the effects of hydrogenic defect diffusion in semiconductors~\cite{storchak1998MuonDiffusionRMP}. Similar $\mu^+$SR experiments have examined Li$^+$ ion diffusion in lithium battery materials~\cite{sugiyama2009LiDiffusion, sugiyama2017LiPowder, umegaki2017LiCathodes}. In superconducting systems, $\mu^+$SR experiments have been used to: measure the superconducting electron density, determine phase diagrams, and characterise vortex lattices~\cite{yaouanc2011MuonTextbook}. $\mu^+$SR experiments have also been applied to biological and chemical systems; for example, to investigate oxygen dependent effects in the radiation treatment of cancer~\cite{pant2014MuonCancer}.\\

Muon spectroscopy is a versatile technique, that has provided insights on a range of physical systems. The technique is still undergoing active development, including the introduction of lower energy muon beams which can be used to probe surface effects~\cite{yaouanc2011MuonTextbook}. However, there are also theoretical challenges for $\mu^+$SR that are desirable to address. While some systems can be analysed using a mean-field, or semi-classical approach, others appear to require a fully quantum treatment~\cite{Celio1986StrongCollisionCorrections, Celio1984ExactMuPolar, holzschuh1984approximate, lord2000StrongCoupledMuExact}. In the following Section, we discuss the simulation and analysis of muon polarisation functions in more detail.\\

\section{Muon polarisation functions}\label{Sec:MuonPolFunctions}
In order to analyse the polarisation function arising from a given $\mu^+$SR experiment, we can compare it to a theoretical polarisation function obtained from a physical model of the studied system. The accuracy of these theoretical polarisation functions is determined by both the level of detail included in the model, and the method used to simulate the model. This can be likened to the field of computational chemistry, where the accuracy of a simulation is determined by both the physical effects included in the model (e.g. the Born-Oppenheimer approximation, or relativistic effects) and the method used to simulate the system (e.g. mean-field approaches, or exact diagonalisation). We focus first on the different methods used to obtain the theoretical polarisation function, before considering other physical effects that can be incorporated into the model.

On a most basic level, one can use a semiclassical, mean-field approach, which considers how the muon polarisation evolves given a model for the surrounding magnetic field distribution. For example, a time-independent magnetic field with Gaussian distributed field strengths leads to the Kubo-Toyabe polarisation function~\cite{kubotoyabe1966book, yaouanc2011MuonTextbook}. This analytically derivable formula, and generalisations of it, are widely applied within the $\mu^+$SR field, and are acceptably accurate in many circumstances.

If greater accuracy is required, techniques have been developed which treat the muon interactions with its local spin environment semi-classically~\cite{dalmasdereotier1992NuclearSpinDynamics, dalmasdereotier1992SemiclassicalTheory}. In some circumstances, these methods can yield high accuracy at a modest computational cost. However, they cannot always account for strong spin-spin interactions, or the effect of quadrupole interactions~\cite{yaouanc2011MuonTextbook}.

Finally, the highest level of accuracy for a given model can be obtained by using a quantum mechanical analysis. The calculations consider a quantum Hamiltonian between the muon and its local spin environment, and evolve the muon polarisation in time according to the Schr\"odinger equation. We will discuss these calculations in more detail below. However, we note here that the cost of these calculations is believed to scale exponentially with the size of the system simulated, due to the computational complexity of storing highly entangled quantum states.\\

In this work, we focus on systems which require a fully quantum treatment in order to obtain accurate polarisation functions. Two techniques have been developed by the muon community for exactly simulating the polarisation function. The first relies on exact diagonalisation of the muon-environment Hamiltonian. Because the thermal energies encountered in $\mu^+$SR experiments are typically much larger than the nuclear energy levels, the environment is normally assumed to be in the maximally mixed state $\rho_e(0) = I_e/D_e$ where $D_e$ is the Hilbert space dimension of the environment. The polarisation function can then be obtained by
\begin{equation}\label{Eq:ExactDiag_1}
\begin{aligned}
        P(t) &= \frac{1}{D_s} \mathrm{Tr}\big{(}\big{[} Z_\mu \otimes I_e \big{]} e^{-iH_st} \big{[} \ket{0}\bra{0}_\mu \otimes I_e \big{]} e^{iH_st}\big{)} \\
        &= \frac{1}{2D_s} \mathrm{Tr}\big{(}\big{[} Z_\mu \otimes I_e \big{]} e^{-iH_st} \big{[} (I_\mu + Z_\mu) \otimes I_e \big{]} e^{iH_st}\big{)} \\
        &= \frac{1}{2D_s} \mathrm{Tr}\big{(}\big{[} Z_\mu \otimes I_e \big{]} e^{-iH_st} \big{[} Z_\mu \otimes I_e \big{]} e^{iH_st}\big{)}.
\end{aligned}
\end{equation}
We then use a resolution of the identity in terms of eigenstates of the system Hamiltonian; $\ket{m}, \ket{n}$
\begin{equation}\label{Eq:ExactDiag_2}
\begin{aligned}
P(t) &= \frac{1}{2D_s} \sum_m \bra{m} Z_\mu e^{-iH_st}  Z_\mu e^{iH_st} \ket{m} \\
&= \frac{1}{2D_s} \sum_m \sum_n \bra{m} Z_\mu e^{-iH_st} \ket{n} \bra{n} Z_\mu e^{iH_st} \ket{m} \\
&= \frac{1}{2D_s} \sum_m \sum_n | \bra{m} Z_\mu \otimes I_e \ket{n} |^2 e^{i(E_m - E_n)t},
\end{aligned}
\end{equation}
where $E_m$ is the eigenvalue of eigenstate $\ket{m}$. As a result, we can see that performing an exact diagonalisation of the system Hamiltonian can be used to calculate the polarisation function. However, the dimension of the system Hamiltonian will scale exponentially with the number of spins considered, which limits the size of these calculations. These exact calculations have been performed for a range of systems~\cite{celio1983ExactDipolarSim,Celio1984ExactMuPolar,holzschuh1984approximate, lord2000StrongCoupledMuExact, yaouanc2011MuonTextbook, lancaster2007molecularmagnets, lancaster2009fluoropolymers,wilkinson2020muonfluorinefull}, with Hilbert space dimensions up to around 2048~\cite{lord2000StrongCoupledMuExact, wilkinson2020muonfluorinefull}. Recently, a method was developed to scale the interactions between the muon and more distant nuclei, in order to act as a proxy for the remaining nuclei in the sample, which showed promising results for muon experiments on CaF$_2$ and NaF~\cite{wilkinson2020muonfluorinefull}. 

An alternative method, developed by \textcite{celio1986trotter}, reduces the memory cost of the calculation, at the expense of introducing statistical uncertainty into the measured result, which can be reduced through sampling. This method uses a first-order product formula based approach, whereby the time evolution operator is divided into a product of operators acting on subsystems of the system (this is also referred to as `Trotterization' by the quantum computing community). When the environment is initially in a pure state $\ket{\phi(0)}_e$, the wavefunction of the system at time $t$ is given by
\begin{equation}\label{Eq:Celio_1}
\begin{aligned}
\ket{\psi_s(t)} &= e^{-iH_st}\ket{0}_\mu \ket{\phi(0)}_e \\
& \approx \bigg{(} \prod_\alpha e^{-iH_s^\alpha t/n} \bigg{)}^n \ket{0}_\mu \ket{\phi(0)}_e \\
& \equiv U_1(t) \big{[} \ket{0}_\mu \ket{\phi(0)}_e \big{]} , 
\end{aligned}
\end{equation}
where $H_s^\alpha$ are subterms in the Hamiltonian (for example, the interaction between a muon and a single nearest-neighbour spin), and $n$ is referred to as the number of Trotter steps used. Equality is recovered in the limit that $n \rightarrow \infty$. \textcite{celio1986trotter} made use of a random-phase approximation inspired method, which sets the wavefunction at time $t$ is as
\begin{equation}\label{Eq:Celio_2}
\begin{aligned}
\ket{\psi_s(t)} &= U_1(t) \bigg{[} \frac{1}{\sqrt{D_e}} \sum_j e^{i\theta_j} \ket{0}_\mu \ket{j}_e \bigg{]},
\end{aligned}
\end{equation}
where $\theta_j$ are randomly chosen on each sample of the algorithm, and $\ket{j}_e$ denote unentangled basis states of the environment. In this case, an approximation for the polarisation function can be obtained from
\begin{equation}\label{Eq:Celio_3}
\begin{aligned}
P(t) &\approx \bra{\psi_s(t)} Z_\mu \otimes I_e \ket{\psi_s(t)} \\
= &\sum_{j} \frac{1}{D_e} \bra{0}_\mu \bra{j}_e U_1(t)^\dag  Z_\mu U_1(t) \ket{0}_\mu \ket{j}_e \\
 &+ \sum_{j,k} \frac{1}{D_e} e^{i(\theta_k - \theta_j)} \bra{0}_\mu \bra{j}_e U_1(t)^\dag  Z_\mu U_1(t) \ket{0}_\mu \ket{k}_e.
\end{aligned}
\end{equation}
The first term is equal to the polarisation function that we wish to measure (up to an error induced by the Trotterization of the time evolution operator). As the phases in the second term are chosen randomly, many of these terms cancel, leading to a small error on the polarisation function. This error can be reduced by repeating the method with independently randomly generated values of $\theta_j$. The error also decreases as the Hilbert space dimension of the system increases. The Trotter error can be reduced by using higher-order product formulae, or by increasing the number of Trotter steps used. We will discuss this source of error in more detail in Sec.~\ref{Sec:QC_for_muons}. While the exact diagonalisation method requires manipulating matrices of dimension $D_s \times D_s$, this product-formula based approach only requires storing the wavefunction, which has dimension $D_s$. This has enabled much larger calculations to be performed using this method, including simulations with Hilbert space dimension $\sim 2^{17}$~\cite{DalmasdeReotier1991CelioMethodStudy} and $2^{26}$~\cite{huang2012MuSupercLargeSim}. This latter calculation is, to the best of our knowledge, the largest $\mu^+$SR calculation performed to date with an exact method. \\

Exact methods for calculating the polarisation function have predominantly been used for two purposes: locating the muon rest site, and studying muon diffusion. Determining the muon rest site is an important challenge in $\mu^+$SR experiments~\cite{blundell2012Bayesian}. If we are able to accurately simulate the muon polarisation function, then the polarisation functions arising from a number of candidate sites can be generated, and compared to the experimental data, in order to determine the most likely muon location. This technique can be employed in conjunction with density functional theory~\cite{bonfa2016MuonDftReview,bernardini2013AbInitioMuons, onuorah2019VibrationalMuons, moller2013HideAndSeek, moller2013MuonFluoridesDFT} and experimental methods~\cite{Luke1991LevelCrossingCopper} to give greater certainty on the muon location. Exact simulation has repeatedly been employed when studying fluorinated materials, due to the strong dipolar interaction between the spin-$\frac{1}{2}$ fluorine nuclei and the muon, and the large electronegativity of the fluorine ion which `traps' the muon~\cite{brewer1986MuonFluorineOriginal}. This effect has been used to locate the muon rest site in fluorinated polymers~\cite{lancaster2009fluoropolymers, nishiyama2003MuonFluorpolymers}, molecular magnets~\cite{lancaster2007molecularmagnets}, and ionic crystals~\cite{wilkinson2020muonfluorinefull, noakes1993MuonFluorineInsulators}. Similar calculations were used to identify the muon rest sites in the high temperature superconductor La$_{2-x}$Sr$_x$CuO$_4$~\cite{huang2012MuSupercLargeSim}. The calculations have often included heuristic terms in the polarisation function, to compensate for the limited system sizes that can be simulated using these costly techniques. 

Polarisation functions calculated from quantum models have also been used to study muon diffusion. As the muon is a light particle, its diffusion between lattice sites is an inherently quantum phenomenon, involving quantum tunnelling through the potential barrier between sites. Calculating the hopping amplitudes from first principles would be an exceedingly costly calculation, requiring an accurate description of the electronic structure of the muon-environment system. Fortunately, diffusion of the muon between lattice sites leads to a measurable change in the polarisation function. We can calculate the muon diffusion rate by comparing the polarisation function obtained in the absence of diffusion (which can be measured at low temperatures) to the polarisation function when diffusion is present. The effect of muon diffusion is typically incorporated into theoretical calculations using the `strong collision model' (SCM). The SCM assumes Markovian, stochastic hopping of the muon between lattice sites. The muon is considered to probabilistically jump to a new site, where the surrounding nuclei are in the maximally mixed state. This leads to a damping of the polarisation function, the strength of which depends on the hopping rate. \textcite{DalmasdeReotier1991CelioMethodStudy} showed that using the mean-field Kubo-Toyabe function in conjunction with the SCM can lead to inaccurate results for the muon hopping strength. More accurate results were obtained by first computing a static polarisation function using the product formula method described above, and then augmenting this with the SCM. \textcite{kadono1989CopperMuonDiffusion,luke1991MuonDiffusionCopper} used the product-formula based method in conjunction with the SCM to analyse the diffusion of muons in copper. They noted that using the Kubo-Toyabe function in conjunction with the SCM would lead to an overestimation of the muon diffusion rate. 

It is known that the strong collision model is of limited accuracy. The assumption that the muon moves to a new site with unpolarised nuclei means that it cannot move to a nearest-neighbour lattice site (as the muon would have altered the polarisation of the nuclei shared between its previous and new sites), or hop back to its previous site. However, quantum tunnelling is a non-Markovian process, and is strongly affected by the presence of phonons in the system. These phonons can lead to a degeneracy between the energies of different sites, increasing the transition probability. In particular, this effect increases the likelihood of the muon returning to its previous position~\cite{yaouanc2011MuonTextbook}. A more accurate model for muon diffusion was developed by \textcite{Celio1986StrongCollisionCorrections}. A rigorous equation for the dynamic polarisation function with a single jump is given by
\begin{equation}\label{Eq:Celio_diffusion}
        \begin{aligned}
        G(t) &= G_0(t)e^{-\nu t} + \nu \int_0^t dt' e^{-\nu t'} F(t, t')\\
        G_0(t) &= \mathrm{Tr}\big{(}e^{-iH_0 t}\rho_s(0) e^{iH_0 t} Z_\mu \big{)} \\
        F(t, t') &= \mathrm{Tr}\big{(} e^{-iH_1 (t-t')} e^{-iH_0 t'} \rho_s(0) e^{iH_0 t'} e^{iH_1 (t-t')} Z_\mu \big{)}
    \end{aligned}
\end{equation}
where $H_0$ is the Hamiltonian at the initial muon site, $H_1$ is the Hamiltonian at the new site, $t'$ denotes the time of the hop, and $\nu$ is the hopping rate. The trace is over all nuclear sites involved in both Hamiltonians. This equation can be generalised to a larger number of jumps. \textcite{Celio1986StrongCollisionCorrections} showed that the inferred hopping rate can differ by around 20\% when using this rigorous approach, compared to using the SCM. To the best of our knowledge, this model has never been employed when analysing experimental data, likely due to the large computational cost of considering many different sites. \\

Having established the scenarios where quantum models are typically utilised in muon spectroscopy analysis, we now consider the interactions present in such systems. The muon-system Hamiltonian typically includes contributions from: dipolar interactions between spins, quadrupole interactions between nuclear spins and electric field gradients in the sample, and the coupling of spins to time-dependent or independent magnetic fields. The dipolar contribution is given by
\begin{equation}\label{Eq:DipoleHamil}
H_D = \frac{1}{2} \sum_{i,j} \frac{\hbar^2 \mu_0 \gamma_i \gamma_j}{4 \pi r_{ij}^3} \bigg{[} \vec{S}_i \cdot \vec{S}_j - 3(\vec{S}_i \cdot \hat{r}_{ij}) (\vec{S}_j \cdot \hat{r}_{ij})   \bigg{]},
\end{equation}
where $\mu_0$ is the permeability of free space, $\gamma_i$ is the gyromagnetic ratio of spin $i$, $\vec{r}_{ij}$ is the vector connecting spins $i$ and $j$, and $\vec{S}_i$ are the vectorized generalised spin matrices for a particle with spin quantum number $s$. We have moved the factor of $\hbar$ from the spin matrices to the coefficient of the sum. For example, for a spin-$\frac{1}{2}$ particle we have $\vec{S}_i = \frac{1}{2}(X_i, Y_i, Z_i)$, where $X_i, Y_i, Z_i$ are the Pauli matrices acting on spin $i$. These can be generalised to spins of higher dimension, which we discuss in more detail in Sec.~\ref{Subsec:MapQubits}. This Hamiltonian contains $O(N^2)$ terms, where $N$ is the number of spins considered in the simulation. 

There is also an interaction between the quadrupole moment of nuclei with $s > \frac{1}{2}$, and any non-zero electric field gradients in the sample. Such electric field gradients can often by induced by the presence of the muon. The quadrupole interaction Hamiltonian is given by~\cite{wilkinson2020muonfluorinefull}
\begin{equation}\label{Eq:QuadHamil_1}
    H_{Q} = \sum_{i \in \mathcal{Q}} \frac{\hbar e Q_i (1 + \gamma_i)}{2S_i(2S_i-1)} \bigg{(} \vec{S_i}^T \cdot \mathbf{G}(\vec{r}_i) \cdot \vec{S_i} \bigg{)},
\end{equation}
where $\mathcal{Q}$ is the set of nuclei in the simulation with quadrupole moments, $e$ is the electron charge, $Q_i$ and $\gamma_i$ are the quadrupole coupling factor and anti-shielding factor (respectively) of the $i$th spin, and $\mathbf{G}(\vec{r}_i)$ is the electric field gradient tensor at position $\vec{r}_i$. The elements of $\mathbf{G}(\vec{r}_i)$ are $G_{\alpha \beta}(\vec{r}_i) = \frac{\partial^2 V(\vec{r}_i)}{\partial r_\alpha \partial r_\beta}$, where $V(\vec{r}_i)$ is the Coulomb potential at position $\vec{r}_i$. This Hamiltonian contains $O(N_{\mathcal{Q}})$ terms, each acting on a single spin (where $N_{\mathcal{Q}}$ is the number of nuclei with quadrupole moments). 

The Zeeman interactions of each spin with an applied magnetic field are given by
\begin{equation}\label{Eq:MagFieldsHamil}
    H_M(t) = \sum_i \hbar \gamma_i \vec{S}_i \cdot \vec{B}(t),
\end{equation}
where $\vec{B}(t)$ is the magnetic field.

Some previous calculations have neglected the dipolar interactions between the nuclear spins, as they are often much smaller than those between the muon and the nuclear spins~\cite{huang2012MuSupercLargeSim,celio1986trotter}. Neglecting these interactions reduces the number of terms in the Hamiltonian to $O(N)$.\\

In this section, we have discussed theoretical methods to generate simulated polarisation functions. The accuracy of these polarisation functions depends on both the model for the system (e.g. whether effects like muon diffusion are taken into account), and the method used to solve the model. The most accurate calculations require a fully quantum model of the system. Unfortunately, the cost of classically simulating the dynamics of a quantum system is expected to scale exponentially with the size of the simulated system. This high cost has restricted the simulation of $\mu^+$SR experiments to small systems, with Hilbert space dimensions of less than $2^{30}$. In Sec.~\ref{Sec:QC_for_muons} we show how this exponential cost can be circumvented by using quantum hardware as the simulation platform.

\section{Quantum computing}\label{Sec:QuantumComputingIntro}
Quantum computing leverages quantum mechanical effects in order to perform certain information processing tasks more efficiently than appears possible with classical computers. Quantum computers were originally proposed as efficient simulators of other quantum systems~\cite{feynman1982simulating}, and were later formalised as a more general computational device~\cite{deutsch1985quantum}. Since these initial proposals, a number of quantum algorithms have been developed which appear to asymptotically outperform known classical algorithms. These include algorithms for factoring numbers~\cite{shor1994algorithms}, searching databases~\cite{grover1996fast}, and simulating quantum systems~\cite{Lloyd1073}. There has also been significant progress in developing platforms on which to run these algorithms. Physical qubits can be created in a number of different systems~\cite{Ladd2010quantumcomputingreview}. These include: using the two lowest energy levels of superconducting circuit resonators~\cite{shnirman1997superconducting,nakamura1999superconducting}, path or polarization degrees of freedom in linear optical photonic systems~\cite{Knill2001linearoptics}, or a pair of energy levels in ions~\cite{cirac1995trappedions,leibfried2003trappedionrmp}. Quantum computing is currently referred to as being in the `noisy, intermediate-scale quantum' (NISQ) era~\cite{preskill2018quantum}. Current quantum computers possess up to around $50$ qubits, and exhibit gate error rates of $\sim 10^{-3}$, at best~\cite{google2019supremacy}. These limited computational resources mean that we have so far been unable to run classically challenging instances of the algorithms referenced above. In this section, we provide an introduction to quantum computing. We refer the interested reader to the textbook by \textcite{nielsen2002quantum} for more information.

In this work, we focus on the circuit model of quantum computation. This formalism abstracts away the physical details of the hardware implementation, and treats the qubits as generic two-level systems. The computational basis states of the qubit Hilbert space are taken to be
\begin{equation}
\ket{0} = 
\begin{bmatrix}
1 \\
0\\
\end{bmatrix},~~
\ket{1} =
\begin{bmatrix}
0 \\
1 \\
\end{bmatrix}.
\end{equation}

A general single qubit state is described by 
\begin{gather}\label{EqQubit}
\ket{\psi} = \alpha \ket{0}+\beta \ket{1}
\end{gather}
where $\alpha, \beta$ are complex amplitudes, and the state is normalised to one. Qubits are manipulated by applying quantum logic gates to the system. These gates are defined by unitary matrices acting on the qubit wavefunction. Typical single qubit gates include the Pauli gates
\begin{equation}
X = 
\begin{bmatrix}
0 & 1 \\
1 & 0 \\
\end{bmatrix},~~
Y = 
\begin{bmatrix}
0 & -i \\
i & 0 \\
\end{bmatrix},~~
Z = 
\begin{bmatrix}
1 & 0 \\
0 & -1 \\
\end{bmatrix},
\end{equation}
the single qubit rotation gates
\textcolor{black}{\begin{align}
R_{x}(\theta) = e^{\left (\frac{{-i \theta X}\,}{2} \right)},~ R_{y}(\theta) = e^{\left (\frac{{-i \theta Y}\,}{2} \right)},~
R_{z}(\theta) = e^{\left (\frac{{-i \theta Z}\,}{2} \right)}
\end{align}}
and the Hadamard and T gates
\begin{equation}
	\mathrm{H} = \frac{1}{\sqrt{2}} 
\begin{bmatrix}
1& 1\\
1& -1
\end{bmatrix},
\quad \mathrm{T} =
\begin{bmatrix}
1& 0\\
0& e^{i\pi / 4}
\end{bmatrix}.
\end{equation}

A system of $Q$ qubits is described by a normalised wavefunction in the Hilbert space of dimension $2^Q$. We can construct wavefunctions that cannot be decomposed into tensor products of individual qubits by using multi-qubit entangling gates, such as the controlled-NOT  (CNOT) gate
\begin{equation}
\begin{bmatrix}
1& 0 & 0 & 0\\
0& 1 & 0 & 0\\
0& 0 & 0 & 1\\
0& 0 & 1 & 0
\end{bmatrix}.
\end{equation}

A quantum circuit consists of a sequence of unitary gates, applied to a well-defined initial state, such as $\ket{\bar{0}} = \ket{0} ... \ket{0}$. Qubits can be measured in the computational basis, either at an intermediate stage of the computation, or at the end of the calculation. When a qubit is measured in the computational basis, it will `collapse' to state $\ket{0}$ or state $\ket{1}$, with probability $|\alpha|^2$ and $|\beta|^2$, respectively. 

While some quantum algorithms (such as Ref.~\cite{grover1996fast}) extract their solution from a single measurement of the qubits, other algorithms require measurements of observables, $O$, which are represented by Hermitian matrices. For algorithms in the latter category, we typically seek the average value of an observable over many measurements, $\bar{O} = \bra{\psi}O\ket{\psi}$. In this work, we will predominantly be concerned with measuring the spin-polarisation of a qubit representing the muon. To measure the polarisation along the $Z$ axis, we can simply prepare the desired state, and measure the qubit representing the muon in the computational basis. We assign a value of +1 to the outcome $\ket{0}$, and $-1$ to the outcome $\ket{1}$. We repeat this process a number of times, and average the results. To measure the polarisation along a different axis we use a single qubit rotation to rotate the measurement axis to the $Z$ axis, and then apply the procedure described above. For example, when measuring $X$, we use the $H$ (Hadamard) gate to rotate the basis. In Sec.~\ref{Subsec:Measurement} we discuss a more complex, but asymptotically more efficient way of measuring the expectation value of Hermitian observables. \\

The formalism introduced above can be used to construct and describe quantum algorithms, by specifying the actions applied to individual qubits. These abstract instructions are then converted into physical controls applied to the quantum hardware, such as applying laser pulses to excite a trapped ion qubit. These physical controls are imperfect, introducing noise to the system. Moreover, the quantum states constructed are sensitive to decoherence caused by interaction with the surrounding environment. These processes introduce noise into the system, which can be modelled in a number of ways. The presence of noise can be overcome by using quantum error correction codes. These codes construct a single logical qubit from highly entangled states of a number of physical qubits. It is important to note the high overheads introduced by quantum error correction; a common rule of thumb is that around $10^3 - 10^4$ physical qubits per logical qubit may be required to solve classically intractable problem instances~\cite{fowler2012surface, PhysRevA.95.032338, campbell2017roads}. We will elaborate more on this resource estimation in Sec.~\ref{Subsec:ResourceEstimates}. We refer the interested reader to Refs.~\cite{terhal2015error,devitt2013quantum, Raussendorf2012error,lidar2013qec} for a more comprehensive discussion of quantum error correction. \\

The limited number of qubits in current NISQ devices precludes our ability to perform quantum error correction. Instead, alternative algorithmic error mitigation strategies have been developed, which effectively seek to identify the noiseless signal from an increased number of noisy experimental repetitions. We refer the interested reader to Ref.~\cite{mcardle2020chemistry} for a more detailed discussion of error mitigation techniques. In this work, we will place particular focus on the extrapolation method of error mitigation. Error extrapolation uses expectation values obtained at multiple different physical noise levels to infer the noiseless expectation value~\cite{Li2017, endo2017practical, temme2017mitigation, giurgica2020extrapolation, cai2020extrapolation}. We can increase the noise level in the quantum processor in a number of ways, including `stretching' the duration of gates~\cite{kandala2018extending}, or by converting the noise into a Pauli channel and then introducing additional Pauli errors with the appropriate probabilities~\cite{Li2017}.

\section{Quantum simulation of $\mu^+$SR}\label{Sec:QC_for_muons}

\begin{figure*}
\includegraphics[width=1.5\columnwidth]{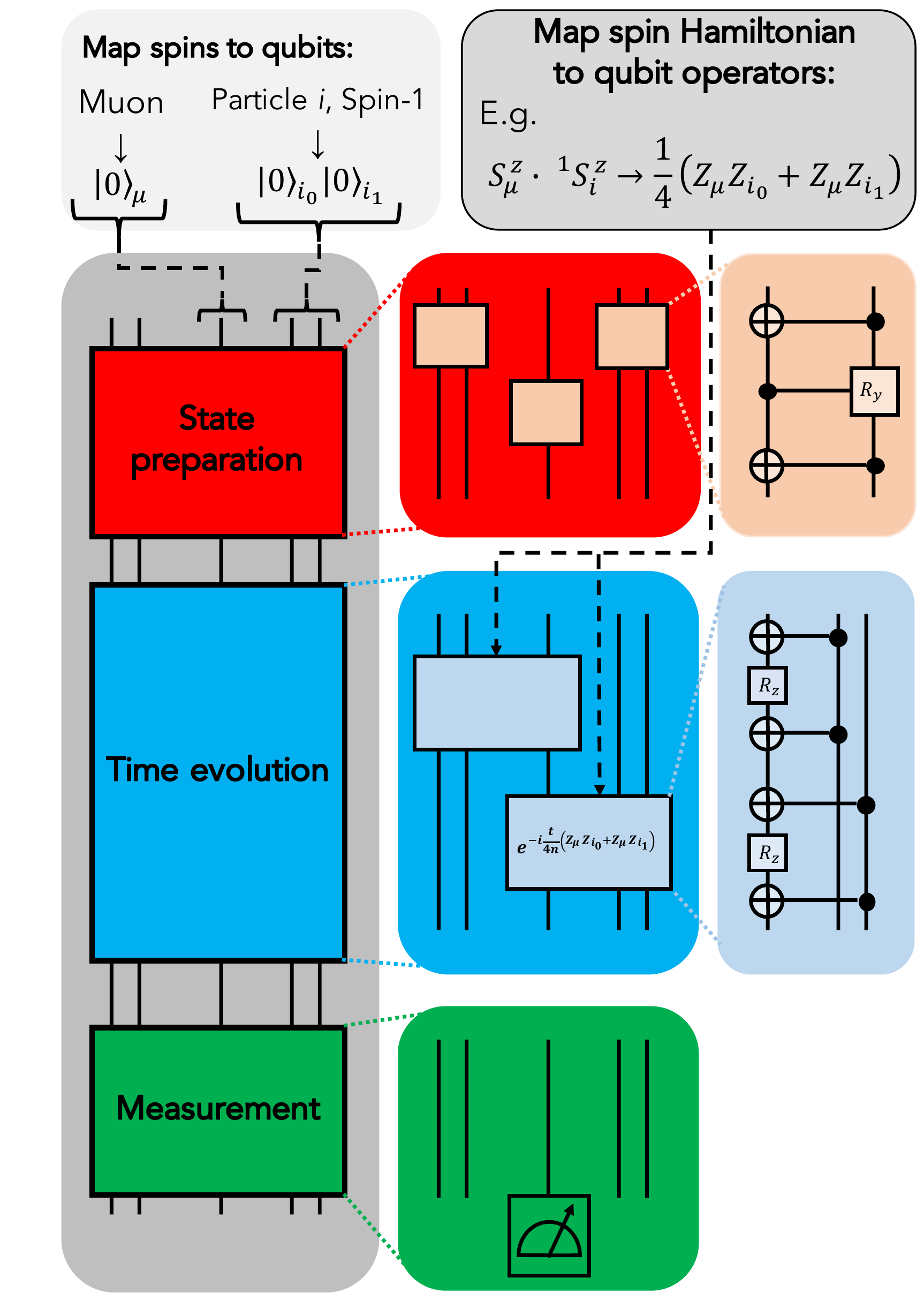}
 \caption{An outline of the proposed algorithm for simulating muon polarisation functions on quantum computers. We first map the spins onto qubits via the mapping in Sec.~\ref{Subsec:MapQubits}, which also enables us to map the spin Hamiltonian to qubit operators. We prepare the desired initial state on these qubit registers using the techniques described in Sec.~\ref{Subsec:PrepInitialState}. We can then evolve this state in time, using the methods in Sec.~\ref{Subsec:TimeEvolution}. We can measure the polarisation value of the muon qubit, as described in Sec.~\ref{Subsec:Measurement}. This diagram shows the most simple form of the algorithm. This circuit must be repeated many times in order to estimate the $Z$ expectation value of the muon at a single time value. We must repeat this procedure for each desired time value. This will generate a simulated polarisation function, which can be incorporated into, for example, an optimisation routine to determine the muon rest site, by fitting the generated data to experimental data.}\label{Fig:Muon_circuit_diagram}
\end{figure*}

In this section, we illustrate how quantum computers can be used to analyse muon spectroscopy data. We show that a quantum computer is able to simulate muon polarisation functions using resources scaling polynomially with the size of the simulated system. This is in constrast to the classical methods for simulating muon polarisation functions described in Sec.~\ref{Sec:MuonPolFunctions}, which required exponentially scaling resources. Our algorithm is illustrated in Fig.~\ref{Fig:Muon_circuit_diagram}, and can be summarised as follows:
\begin{enumerate}
    \item Map the spin system of interest to qubits.
    \item Prepare the quantum registers in the desired initial state.
    \item Evolve the system in time for the desired duration, $t$.
    \item Measure the muon $Z$ expectation value, $P(t)$.
\end{enumerate}
We can repeat steps $2-4$ of this process at many different $t$ values in order to obtain a simulated version of the polarisation function for the given system. Below, we discuss in more detail how each of these steps can be implemented. \\

\subsection{Mapping the spin system to qubits}\label{Subsec:MapQubits}
We first need to map the spin system under consideration onto the quantum computer, such that the spin states of the system correspond to valid states of the quantum register. For the case of spin-$\frac{1}{2}$ particles, this mapping takes a simple form 
\begin{equation}\label{Eq:Spin_half_map}
\begin{aligned}
    &\bigg{|} \frac{\textbf{1}}{\textbf{2}} \bigg{\rangle} := \ket{0} \\ 
    &\bigg{|} \textbf{--} \frac{\textbf{1}}{\textbf{2}} \bigg{\rangle} := \ket{1}.
\end{aligned}
\end{equation}
This leads to a natural mapping of the spin operators of particle $i$
\begin{equation}\label{Eq:Spin_half_ops}
\begin{aligned}
    &S_i^x = \frac{1}{2} X_i \\ &S_i^y = \frac{1}{2} Y_i \\
    &S_i^z = \frac{1}{2} Z_i.
\end{aligned}
\end{equation}
When considering particles with higher spin $s$, there are a number of possible mappings that we could use. One approach is to store the spin of the particle in a register with $\mathrm{log_2}(2s+1)$ qubits~\cite{Sawaya2020digitalvibrational}. However, this compact mapping can lead to more complicated expressions for the spin operators, and may be problematic if the spin multiplicity is not a power of $2$. In this work, we employ a spin-to-qubit mapping based on symmetric quantum states of a given Hamming weight (known as Dicke states)~\cite{koczor2019thesis}. For example, a spin-1 particle is stored as
\begin{equation}\label{Eq:Spin_one_map}
\begin{aligned}
    &\ket{\textbf{1}} := \ket{00} \\ &\ket{\textbf{0}} := \frac{1}{\sqrt{2}}\big{(}\ket{01} + \ket{10}\big{)} \\
    &\ket{\textbf{--1}} := \ket{11}.
\end{aligned}
\end{equation}
This mapping can be derived by considering the joint Hilbert space of two spin-$\frac{1}{2}$ particles, which can be partitioned into a spin-1 triplet space, and a spin-0 singlet space. This mapping is an embedding of the 3-dimensional triplet space into the 4-dimensional two-qubit Hilbert space. The spin operators for particle $i$ are given by
\begin{equation}\label{Eq:Spin_one_ops}
\begin{aligned}
    &S_i^x = \frac{1}{2} \big{(} X_{i_0} + X_{i_1} \big{)} \\ &S_i^y = \frac{1}{2} \big{(} Y_{i_0} + Y_{i_1} \big{)}\\
    &S_i^z = \frac{1}{2}\big{(} Z_{i_0} + Z_{i_1} \big{)}.
\end{aligned}
\end{equation}
We can generalise this mapping for a spin-$s$ particle by assigning each of the states $\ket{j}, j\in [-s, ..., s]$ to the symmetric superposition of states with Hamming weight $h$ such that $h=s-j$. For example, a spin-$\frac{3}{2}$ particle can be represented by  
\begin{equation}\label{Eq:Spin_3_2_map}
\begin{aligned}
    & \bigg{|} \frac{\textbf{3}}{\textbf{2}} \bigg{\rangle} := \ket{000} \\
    & \bigg{|} \frac{\textbf{1}}{\textbf{2}} \bigg{\rangle} := \frac{1}{\sqrt{3}} \big{(} \ket{001} + \ket{010} + \ket{100} \big{)} \\
    & \bigg{|} \textbf{--} \frac{\textbf{1}}{\textbf{2}} \bigg{\rangle} := \frac{1}{\sqrt{3}} \big{(} \ket{011} + \ket{101} + \ket{110} \big{)} \\
    & \bigg{|} \textbf{--} \frac{\textbf{3}}{\textbf{2}} \bigg{\rangle} := \ket{111}.
\end{aligned} 
\end{equation}
The spin operator for a spin-$s$ particle $i$ is given by
\begin{equation}
    {}^sS_i^\alpha = \frac{1}{2} \sum_{j=0}^{2s-1} P_{i_j}^\alpha, 
\end{equation}
where $\alpha \in [x,y,z]$ denotes which of the Pauli matrices $P^\alpha \in [X, Y, Z]$ are used, and $j$ denotes the qubits in register $i$. For example, the $S^x$ operator on particle $i=1$ with spin $s=\frac{3}{2}$ is given by
\begin{equation}
    {}^{\frac{3}{2}}S_1^x = \frac{1}{2} \big{(} X_{1_0} + X_{1_1} + X_{1_2} \big{)}.
\end{equation}

The total number of qubits required is given by $\sum_{i=1}^N 2s_i$, where $s_i$ is the spin of the $i$th particle, and $N$ is the number of particles in the system. For example, for six spin-$\frac{3}{2}$ particles, plus a muon, we need $\big{(}6 \times 2 \times \frac{3}{2} \big{)} + \big{(}1 \times 2 \times \frac{1}{2} \big{)} = 19$ qubits. We can use this mapping to obtain qubit representations of the system Hamiltonians given in Eqs.~(\ref{Eq:DipoleHamil}, \ref{Eq:QuadHamil_1}, \ref{Eq:MagFieldsHamil}). The dipole Hamiltonian is the dominant contribution to the number of terms in the Hamiltonian. Each term ${}^sS_i^\alpha \otimes {}^{s'}S_j^\beta$ is mapped to
$4ss'$ two-qubit Pauli terms. As a result, the Hamiltonian contains up to $\mathcal{O}(N^2 s_{\mathrm{max}}^2)$ two-qubit Pauli terms, where $s_{\mathrm{max}}$ is the largest spin value in the system. \\

\subsection{Preparing the initial state of the system}\label{Subsec:PrepInitialState}
As discussed in Sec.~\ref{Sec:MuonPolFunctions}, the initial state of the system is typically taken as $\rho_0 = \ket{0}\bra{0}_\mu \otimes \frac{I_e}{D_e}$. There are two routes to effectively time evolve this mixed initial state. The first approach is to emulate Nature; we can prepare the environment register in a state $\ket{k}_e = \bigotimes_{i \in e} \ket{j_i}_i$ that is the tensor product of each environment spin in a randomly selected $s_z$ basis state. Each state $\ket{k}_e$ is chosen with probability $p_k = \frac{1}{D_e}$. By repeating the simulation many times, we obtain
\begin{equation}\label{Eq:SimpleSample}
\begin{aligned}
    \sum_k p_k \mathrm{Tr} \bigg{(} (Z_\mu \otimes I_e) e^{-iHt} (\ket{0}\bra{0}_\mu \otimes \ket{k}\bra{k}_e) e^{iHt} \bigg{)} \\
    = \mathrm{Tr} \bigg{(} (Z_\mu \otimes I_e) e^{-iHt} (\ket{0}\bra{0}_\mu \otimes \frac{I_e}{D_e}) e^{iHt} \bigg{)},
\end{aligned}
\end{equation}
as required. This method converges as $1/\sqrt{\omega}$, where $\omega$ is the number of samples taken.

The second approach to effectively sample from the time evolved initial mixed state is a modified version of the random-phase-approximation method~\cite{celio1986trotter} discussed in Eqs.~(\ref{Eq:Celio_2}--\ref{Eq:Celio_3}). We first initialise the environment register in an equal superposition of all possible $\ket{k}_e$ states. For the case of spin-$\frac{1}{2}$ particles, this can be accomplished by applying a Hadamard gate to each of the qubits. For the case of higher-spin particles, we will discuss below how to construct this superposition. These steps will prepare the state
\begin{equation}\label{Eq:QuantumCelio}
    \ket{\psi} = \frac{1}{\sqrt{D_e}} \sum_k \ket{0}_\mu \ket{k}_e,
\end{equation}
which can be compared with the state given in Eq.~(\ref{Eq:Celio_2}). In order to generate the desired random phase for each basis state $\ket{k}_e$, we apply random single particle $R_z(\theta)$ rotations to each particle. This procedure will not generate a state with completely independent phases for each basis state. If it is necessary to further randomise the state, we can apply layers of $R_z$ and controlled-$R_z$ rotations between the different particles. We average the results of several simulations (each with a different set of randomly chosen $\theta_i$) to obtain the polarisation function, as shown by Eq.~(\ref{Eq:Celio_3}). While the asymptotic convergence properties of the two methods are the same, we expect the latter method to yield a smaller error for a given number of samples.\\

For both of the methods discussed above, we must construct the states $\ket{k}_e$, either alone, or in superposition. These states are tensor products of the environment spins each in a arbitrary state $\ket{j}$, that is an $s_z$ eigenstate. As discussed in Sec.~\ref{Subsec:MapQubits}, we have mapped these spin states onto qubit Dicke states. Efficiently constructing Dicke states (or superpositions of Dicke states) on quantum computers has remained challenging for a number of years~\cite{bacon2005schur, bacon2006schur, kirby2018schur, Krovi2019schur}, but has recently been made possible with the elegant inductive solution of \textcite{bartschi2019dicke}. Their algorithm can be summarised as follows, and is explained in more detail in Appendix.~\ref{Appendix:Dicke_States}. We first define a Dicke state with Hamming weight $h$, on $n$ qubits, as $\ket{D^n_h}$. For example, the $\ket{\textbf{--1/2}}$ state in Eq.~(\ref{Eq:Spin_3_2_map}) equates to $\ket{D^3_2}$. We can then observe that
\begin{equation}\label{Eq:DickeStateRelation}
    \begin{aligned}
    \ket{D^n_h} = \sqrt{\frac{h}{n}}\ket{D^{n-1}_{h-1}}\otimes \ket{1} + \sqrt{\frac{n-h}{n}}\ket{D^{n-1}_h}\otimes \ket{0}.
    \end{aligned}
\end{equation}
We assume the existence of a unitary operator $U_{n,k}$ such that $U_{n,k} \ket{0}^{\otimes n-h} \ket{1}^{\otimes h} = \ket{D^n_h}$ for all $h \leq k$. Through induction, it can be shown that this unitary operator exists, and how to construct it from typical single- and two-qubit gates. For example, note that
\begin{equation}\label{Eq:DickeCircuitEq1}
    \begin{aligned}
    \ket{D^n_h} = U_{n,k} \ket{0}^{\otimes n-h} \ket{1}^{\otimes h},
    \end{aligned}
\end{equation}
and
\begin{equation}\label{Eq:DickeCircuitEq2}
    \begin{aligned}
    \ket{D^n_h} =& \sqrt{\frac{h}{n}}\ket{D^{n-1}_{h-1}}\otimes \ket{1} + \sqrt{\frac{n-h}{n}}\ket{D^{n-1}_h}\otimes \ket{0}. \\
    =& U_{n-1, k} \bigg{[} \sqrt{\frac{h}{n}} \ket{0}^{\otimes n-h} \ket{1}^{\otimes h} \\
    &+ \sqrt{\frac{n-h}{n}} \ket{0}^{\otimes n-1-h} \ket{1}^{\otimes h} \ket{0}  \bigg{]} \\
    =& U_{n-1, k} V_{n,k} \ket{0}^{\otimes n-h} \ket{1}^{\otimes h},
    \end{aligned}
\end{equation}
implying that
\begin{equation}\label{Eq:DickeCircuitEq3}
    U_{n,k} = U_{n-1, k} V_{n,k}.
\end{equation}
As shown in Appendix.~\ref{Appendix:Dicke_States}, these relations can be repeated recursively, to obtain a circuit purely in terms of the $V_{n,k}$ type gates. \textcite{bartschi2019dicke} show how these gates can be constructed from CNOT gates and single qubit rotations (see also Appendix.~\ref{Appendix:Dicke_States}). The entire state preparation circuit has a depth of $\mathcal{O}(n)$, and requires $\mathcal{O}(kn)$ gates, even on a 1D linear chain of qubits. The algorithm also requires no additional ancillary qubits. Because the unitary $U_{n,k}$ that creates the Dicke states was defined to work for all input states with $h \leq k$, we can use unitary $U_{n,n}$ create all of the Dicke states $\ket{D^n_{h \leq n}}$. As a result, if we input the superposition
\begin{equation}\label{Eq:GenerateDickeSuper}
    \sqrt{\frac{1}{n+1}} \sum_{k=0}^{n} \ket{0}^{\otimes n-k} \ket{1}^{\otimes k}
\end{equation}
then we obtain an equally weighted superposition of all of the possible Dicke states. If we apply this approach to each environment spin, then this will generate the state shown in Eq.~(\ref{Eq:QuantumCelio}). It is straightforward to generate the state in Eq.~(\ref{Eq:GenerateDickeSuper}), using a ladder of control-R$_y$ gates, as shown in Ref.~\cite{bartschi2019dicke}. As a result, a similar number of gates are required for both of the initial state preparation methods described in this section.

\subsection{Evolving the state in time}\label{Subsec:TimeEvolution}
Once we have generated the desired initial state, we can then evolve it in time for the desired duration. There are a number of approaches that can be used for performing time evolution on quantum computers~\cite{low2016hamiltonian, PhysRevLett.118.010501, PhysRevLett.114.090502, Li2017, campbell2018random}. In this work, we focus on product formula based approaches (also referred to as `Trotterization'), which have been observed to lead to lower gate counts than some asymptotically more efficient algorithms when considering the time evolution of spin systems~\cite{childs2017toward}. Trotter-based methods implement time evolution by dividing the propagator into a product of time evolutions under subterms in the Hamiltonian, that have known decompositions into single and two-qubit gates. Detailed derivations of the error resulting from Trotterization are given in Ref.~\cite{childs2019trottererror}, and we report some of that work's key results below. We decompose the Hamiltonian into a sum of Pauli strings, as $H = \sum_\alpha h_\alpha H_\alpha$, where $h_\alpha$ are real coefficients. A first-order Trotter decomposition of the time evolution operator is given by
\begin{equation}\label{Eq:FirstOrderTrotter}
    U_1(t) := \bigg{(} \prod_j e^{-\frac{it}{n}h_j H_j} \bigg{)}^n,
\end{equation}
where $n$ is the number of Trotter steps used. The error in the first-order product formula is upper bounded by
\begin{equation}\label{Eq:FirstOrderTrotterLoose}
    || U(t) - U_1(t) || := \epsilon \sim \mathcal{O}\bigg{(}\frac{(L \Lambda t)^2}{n} \bigg{)}
\end{equation}
where $U(t)$ is the true time evolution operator, $||~||$ denotes the spectral norm,  $L$ is the number of terms in $H$, and $\Lambda = \mathrm{max}_\alpha ||h_\alpha||$. A tighter bound on the error due to first-order Trotterization is given by
\begin{equation}\label{Eq:FirstOrderTrotterTight}
    \epsilon \sim \frac{t^2}{2n} \sum_{i} \bigg{|} \bigg{|} \sum_{j>i} [h_j H_j, h_i H_i]\bigg{|} \bigg{|},
\end{equation}
which is known to be tight, up to an application of the triangle inequality. We can also consider higher-order product formulae, which yield improved error-scaling. The second-order product formula is given by
\begin{equation}\label{Eq:SecondOrderTrotter}
    U_2(t) := \bigg{(} \prod_{j=1}^L e^{-\frac{it}{2n}h_j H_j} \prod_{k=L}^1 e^{-\frac{it}{2n}h_k H_k} \bigg{)}^n.
\end{equation}
An upper-bound on the second-order trotter error is given by 
\begin{equation}\label{Eq:SecondOrderTrotterLoose}
    || U(t) - U_2(t) || := \epsilon \sim \mathcal{O}\bigg{(}\frac{(L \Lambda t)^3}{n^2} \bigg{)},
\end{equation}
and a tighter bound (tight up to an application of the triangle inequality) is given by
\begin{equation}\label{Eq:SecondOrderTrotterTight}
\begin{aligned}
    \epsilon \sim \frac{t^3}{n^2} \bigg{[} &\frac{1}{12} \bigg{(}\sum_{i} \bigg{|} \bigg{|} \sum_{j>i} \sum_{k>i} [h_k H_k, [h_j H_j, h_i H_i]] \bigg{|} \bigg{|} \bigg{)} \\
    + &\frac{1}{24} \bigg{(}\sum_{i} \bigg{|} \bigg{|} \sum_{j>i} [h_i H_i, [h_i H_i, h_j H_j]] \bigg{|} \bigg{|} \bigg{)}
    \bigg{]}.
    \end{aligned}
\end{equation}
It has also been shown that introducing aspects of randomized compilation into these algorithms can lower the gate counts required to obtain results of a given accuracy. For example, it has been shown that randomly permuting the ordering of the product formula terms in each step can reduce the error scaling obtained~\cite{childs2018faster}. An alternative randomized procedure, known as qDRIFT~\cite{campbell2018random}, probabilistically selects a number of terms from the Hamiltonian according to their strength, and then evolves under these terms (each for an equal duration in time). A single step of qDRIFT is equivalent to implementing the channel
\begin{equation}\label{Eq:Qdrift}
    \mathcal{E}(\rho) = \sum_j \frac{h_j}{\lambda} e^{-i\frac{\lambda t}{N} H_j} \rho e^{i\frac{\lambda t}{N} H_j},
\end{equation}
where $\lambda = \sum_j h_j$ is the 1-norm of the Hamiltonian (when applying qDRIFT, we shift the signs from the $h_j$ coefficients to the $H_j$ operators, such that $h_j$ are all real and positive), and $N$ is the number of qDRIFT steps used in this simulation. The error scaling of qDRIFT is given by
\begin{equation}
    \epsilon \sim \frac{\lambda^2 t^2}{N}.
\end{equation}
Most notably, the error scaling of qDRIFT is independent of the number of terms in the Hamiltonian, making it an interesting candidate for systems with a large number of weakly interacting terms. Given the rapid power-law decay of the dipolar interactions studied in this work, qDRIFT may be an interesting candidate for simulating muon spectroscopy on quantum computers. While a thorough numerical comparison of these approaches is beyond the scope of this work, it would be an interesting area for future study - especially if methods that interpolate between Trotterization and qDRIFT are also considered~\cite{ouyang2019stochastichamiltonian}. 

The aim of our simulation algorithm is to obtain an accurate estimate for the value of the muon polarisation function at a given time. As a result, we are not interested in the Trotter error $\epsilon$ directly, but in the error it induces on $\mathrm{Tr}\big{(}Z_\mu \rho(t)\big{)}$. We expect that the Trotter error will provide a loose upper bound for this error, as has been seen previously~\cite{heyl2019quantum, childs2019trottererror}. \\

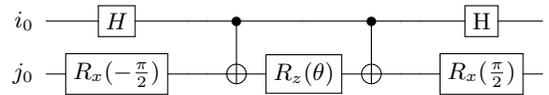
\begin{figure}[hbt]
\centering
\begin{gather*}
\Qcircuit @C=0.8em @R=.7em {
\lstick{i_0}&\gate{H}& \qw & \qw &	\ctrl{1} 	&\qw &	\ctrl{1} & \qw & \qw & \gate{\mathrm{H}} &\qw	\\
\lstick{j_0}&\gate{R_x(-\frac{\pi}{2})}&	\qw &\qw & \targ & \gate{R_z(\theta)} & \targ &	\qw &\qw & \gate{R_x(\frac{\pi}{2})} & \qw	\\
}\\ \\
\end{gather*}
\caption{A quantum circuit implementing the term $\mathrm{exp}(-i\frac{\theta}{2}X_{i_0} Y_{j_0})$.} \label{Fig:TimeEvolution}
\end{figure}

Each of the operators of the form $\mathrm{exp}(-i h_\alpha H_\alpha t)$ can be decomposed into a sequence of single- and two-qubit gates. We illustrate this by considering a term in the dipolar Hamiltonian for spin-$1$ particles. A single term, such as ${}^1S_i^x \otimes {}^{1}S_j^y$ can be mapped to 
\begin{equation}\label{Eq:TrotterExample}
\begin{aligned}
    & e^{-i\theta~{}^1S_i^x \otimes {}^{1}S_j^y} = e^{-i\frac{\theta}{2}(X_{i_0} + X_{i_1}) \otimes (Y_{j_0} + Y_{j_1})} \\ 
    &= e^{-i\frac{\theta}{2}X_{i_0} Y_{j_0}} e^{-i\frac{\theta}{2}X_{i_0} Y_{j_1}} e^{-i\frac{\theta}{2}X_{i_1} Y_{j_0}} e^{-i\frac{\theta}{2}X_{i_1} Y_{j_1}}.
\end{aligned}
\end{equation}
There is no Trotter error arising from this decomposition, as all of the terms in the exponential commute with each other. In Fig.~\ref{Fig:TimeEvolution} we show a quantum circuit that performs time evolution under one of the exponentials in this term.

As discussed above, the dipolar Hamiltonian can be decomposed into up to $L \sim \mathcal{O}(N^2 s_{\mathrm{max}}^2)$ two-qubit Pauli terms, where $s_{\mathrm{max}}$ is the largest spin value in the system. Considering a second-order product formula, the number of Trotter steps required to obtain Trotter-error $\epsilon$ is upper bounded by
\begin{equation}\label{Eq:2ndOrderSystemSpec}
    n \sim \frac{N^3 s_{\mathrm{max}}^3 \Lambda^\frac{3}{2} t^\frac{3}{2}  }{\epsilon^\frac{1}{2}}.
\end{equation}
Each second-order Trotter step requires $ \mathcal{O}(L)$ gates to implement, resulting in a total gate count that scales at worst as
\begin{equation}\label{Eq:2ndOrderGateCount}
    G \sim \frac{N^5 s_{\mathrm{max}}^5 \Lambda^\frac{3}{2} t^\frac{3}{2}  }{\epsilon^\frac{1}{2}}.
\end{equation}
We note that this scaling is obtained without considering any possible compilations, and by using the loose second-order Trotter error bounds. As a result, it is likely that this scaling estimate could be tightened significantly when considering a real system of interest. Similar results were obtained in the context of quantum chemistry calculations. While initial estimates suggested a scaling of $\mathcal{O}(N^{11})$~\cite{wecker2014gates}, more careful analysis reduced the asymptotic scaling to $\mathcal{O}(N^{5.5})$~\cite{babbush2015chemical}. \\

\subsection{Measuring the polarisation}\label{Subsec:Measurement}
The final stage of the algorithm consists of measuring the $Z$ expectation value of the muon qubit. The most straightforward way to measure this value is described in Sec.~\ref{Sec:QuantumComputingIntro}. We repeatedly prepare the desired state at time $t$, measure the muon qubit in the computational basis, and average the results. The standard error in the estimate obtained is given by
\begin{equation}\label{Eq:StandardErrorShot}
    \alpha = \frac{\sqrt{1-\bar{Z}_\mu^2}}{\sqrt{M}},
\end{equation}
where $M$ is the number of samples taken. In order to obtain an error rate of $10^{-3}$ with this approach, we would require up to $10^6$ samples. The repetition rate of a quantum processor can depend on a number of factors, including the depth of the circuit, and the speed of executing quantum gates. The speed of implementing logic gates in a quantum computer depends on the hardware considered, and can range from $ 10 - 100$~MHz in superconducting qubits~\cite{wendin2017superconducting} to $10$~KHz -- $1$~MHz in trapped ion qubits~\cite{schafer2018fast}. As a result, even if a gate depth of only $10^3$ was required for the circuit, obtaining an estimate of $\bar{Z}_\mu$ to a precision of $10^{-3}$ would take at least 10 seconds to calculate on a superconducting qubit processor (in this estimate we have neglected the time taken for qubit readout and initialisation, which can often be longer than the gate time in superconducting qubits~\cite{wendin2017superconducting}). These estimates become even more costly if the overhead of quantum error correction is taken into account. When performing quantum error correction, the logical gate speed of the quantum computer depends on the time taken to measure and decode the error syndromes of all of the physical qubits. This has previously been assumed to be on the order of \SI{10}{\micro\second}~\cite{kivlichan2019condensedtrotter}, which would lead to an estimate of around $2.8$~hours to measure $\bar{Z}_\mu$ to a precision of $10^{-3}$ using the method described above. This is too slow for our purpose of analysing polarisation curves consisting of hundreds of data points. While this direct sampling method has a time cost of $\mathcal{O}(1/\epsilon^2)$, there are alternative quantum algorithms which can reduce the time cost to $\mathcal{O}(1/\epsilon)$. These techniques rely on a combination of quantum amplitude amplification and phase estimation~\cite{wang2018generalisedvqe, knill2007parallelphaseest}. These approaches use a constant number of samples, but require a circuit depth of $\mathcal{O}(d_U/\epsilon)$, where $d_U$ is the circuit depth required to implement the time evolution circuit $U(t)$. This is achieved by repeatedly evolving the register under a controlled version of the operator
\begin{equation}
    \begin{aligned}
    \Omega &:= \Omega_1 \Omega_2, \\
    \Omega_1 &= U(t) \big{[} I_{e,\mu} - 2\ket{\Phi}\bra{\Phi} \big{]} U^\dag (t), \\
    \Omega_2 &= Z_\mu \Omega_1 Z_\mu,
    \end{aligned}
\end{equation}
where $\ket{\Phi}$ is the initial state of the system. This conditional evolution is controlled by the state of an auxiliary register. Controlling the evolution on an auxiliary register enables us to perform quantum phase estimation on the unitary $\Omega$, which we can use to extract the value of $\bar{Z}_\mu$ to the desired precision. A more detailed discussion of this approach is given in Refs.~\cite{wang2018generalisedvqe,knill2007parallelphaseest}. \\

The steps outlined in this section can be used to obtain a simulated polarisation function for the system of interest. In order to use this function to analyse experimental data, the quantum simulation routine can be incorporated into an optimisation loop. As an example, we consider the problem of locating the muon rest site. We first specify the positions of each particle in the system, and use this information to generate the spin and qubit Hamiltonians. We can then use the quantum simulation routine outlined above to calculate the simulated polarisation function. We can quantify the agreement between the experimental polarisation data and the simulated polarisation function using a suitable loss function, such as $\chi^2$ (the sum of the normalised squared residuals). An optimisation loop can then be used to minimise the value of the loss function, by updating the positions of the particles in the system.



\section{Results}\label{Sec:Results}
In order to investigate the practicality of the algorithm discussed in Sec.~\ref{Sec:QC_for_muons}, we performed numerical simulations of systems with up to 29 qubits. We focus on the dipolar interactions between an implanted muon, and spin-$\frac{1}{2}$ fluorine nuclei in a sample of interest. As discussed in Sec.~\ref{Sec:MuonPolFunctions}, the electronegativity of the fluorine ion (F$^-$) acts as a `trap' for the positively charged muon. We can use the `fingerprint' left by the F$^-$--$\mu^+$ dipolar interaction on the polarisation function to determine the muon rest site. This technique has been used to locate the muon in a range of systems, including the ionic crystals considered in this work~\cite{nishiyama2003MuonFluorpolymers,wilkinson2020muonfluorinefull}. We investigate applying our quantum algorithm to analyse the spectra arising from $\mu^+$SR experiments on the ionic crystal calcium fluoride (CaF$_2$), obtained in Refs.~\cite{wilkinson2020muonfluorinefull},\footnote{The experimental data analysed in this work was kindly provided by S.~Blundell and J.~Wilkinson. The data was taken using the pulsed muon beam at the ISIS Facility, Rutherford Appleton Laboratory, UK. Experiments were performed in zero magnetic field, at a temperature of 50~K. Around 181 million muon decay events are included in the dataset.}. CaF$_2$ provides an ideal test-system for the methods introduced in this work for two main reasons. Firstly, the calcium nuclei have spin-$0$ with an abundance of $\sim99.9$~\%, so we only need to consider the dipolar interaction between the muon and the spin-$\frac{1}{2}$ fluorine nuclei~\cite{wilkinson2020muonfluorinefull}. Secondly, the recent exact simulation results of \textcite{wilkinson2020muonfluorinefull} for this system provide a useful benchmark for our results.

\begin{figure}
\includegraphics[width=0.9\columnwidth]{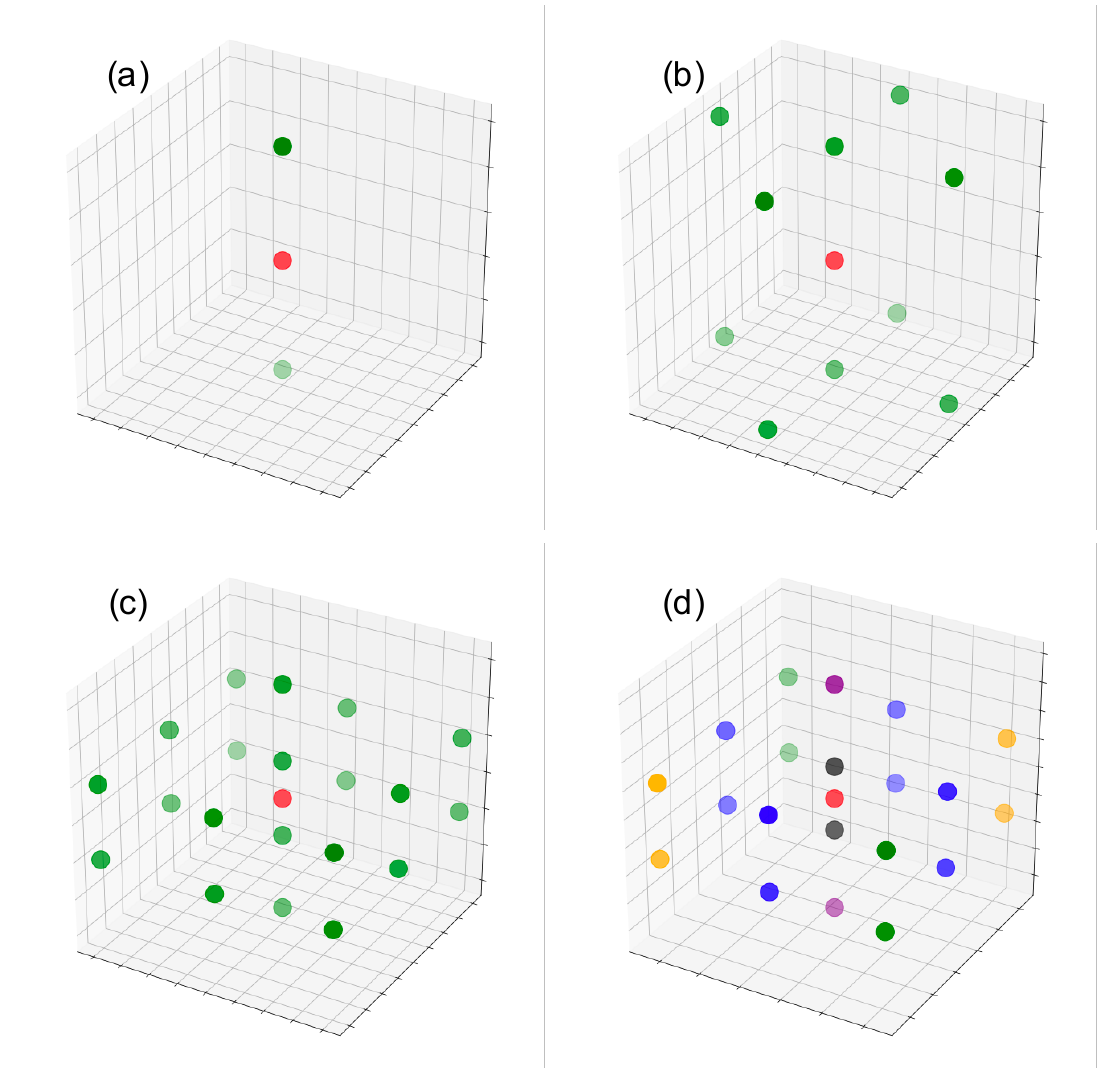}
 \caption{A selection of geometries investigated for the CaF$_2 + \mu^+$ system considered in this work. As discussed in the main text, the calcium ions are spin-0 with an abundance of around $99.9$~\%, so we only include the fluorine ions (green in a - c) and the muon (red) in our simulations. a) The three spin F-$\mu^+$-F system. b) The 11 spin F-$\mu^+$-F system with 8 next-nearest-neighbour fluorines. c) The 21 spin system, which includes the next-nearest 10 fluorines to the muon. d) The 21 spin system, with fluorines grouped by colour. The distances between the muon and each fluorine in a given group are the same. When fitting simulated data for the 21 qubit system to experimental data, we separately parameterised the F-$\mu$ distances according to these groups.} \label{Fig:System_Geometry}
\end{figure}

The geometry studied consists of a simple cubic lattice of F$^-$ ions, with a lattice constant of $2.72$~\AA. The muon implantation site was taken to be between two adjacent fluorines, as shown in Fig.~\ref{Fig:System_Geometry}. As the system is composed of only spin-$\frac{1}{2}$ particles, we can map each particle to a single qubit, as described in Sec.~\ref{Subsec:MapQubits}. We consider a dipolar interaction between all particles; between the muon and the fluorines, and between the fluorines themselves. The Hamiltonian is given by
\begin{equation}\label{Eq:DipoleHamilRepeat}
H_D = \frac{1}{2} \sum_{i,j} \frac{\hbar^2 \mu_0 \gamma_i \gamma_j}{4 \pi r_{ij}^2} \bigg{[} \vec{S}_i \cdot \vec{S}_j - 3(\vec{S}_i \cdot \hat{r}_{ij}) (\vec{S}_j \cdot \hat{r}_{ij})   \bigg{]},
\end{equation}
where $\mu_0$ is the permeability of free space, $\gamma_i$ is the gyromagnetic ratio of spin $i$ ($\gamma_\mu = 2\pi \times 1.355 \times 10^8$ Hz$\cdot$T$^{-1}$,  $\gamma_F = 2\pi \times 4.006 \times 10^7$ Hz$\cdot$T$^{-1}$),  $\vec{r}_{ij}$ is the vector connecting spins $i$ and $j$, and $\vec{S}_i = \frac{1}{2}(X_i, Y_i, Z_i)$, where $X_i, Y_i, Z_i$ are the Pauli matrices acting on qubit $i$. When calculating the polarisation function for this polycrystaline system, we must perform an angular average. This can be obtained using
\begin{equation}\label{Eq:AngularAverage}
\begin{aligned}
    \langle P(t) \rangle = \frac{1}{3} \bigg{[} &\mathrm{Tr}\big{(}Z_\mu e^{-iH_st} \big{[} \ket{0}\bra{0}_\mu \otimes \rho_e(0)\big{]} e^{iH_st}\big{)} \\
    + &\mathrm{Tr}\big{(}X_\mu e^{-iH_st} \big{[} \ket{+}\bra{+}_\mu \otimes \rho_e(0)\big{]} e^{iH_st}\big{)} \\
    + &\mathrm{Tr}\big{(}Y_\mu e^{-iH_st} \big{[} \ket{+Y}\bra{+Y}_\mu \otimes \rho_e(0)\big{]} e^{iH_st}\big{)}\bigg{]} 
\end{aligned}
\end{equation}
where $\ket{+} = \frac{1}{\sqrt{2}}\big{(}\ket{0} + \ket{1}\big{)}$, $\ket{+Y} = \frac{1}{\sqrt{2}}\big{(}\ket{0} + i\ket{1}\big{)}$, are the $+1$ eigenstates of $X$ and $Y$, respectively. We can obtain a more accurate description of the CaF$_2 + \mu^+$ system by considering an increasing number of nearby fluorines. The smallest system considered requires three qubits, representing the muon and its two nearest-neighbour (nn) fluorines. Additional fluorine nuclei are then added in `shells' determined by their distance from the muon. In the `next-nearest-neighbour' (n-nn) shell, there are an additional 8 nuclei. The `next-next-nearest-neighbour' (nn-nn) shell contributes an additional 10 nuclei. The nnn-nn shell adds 8 nuclei. As a result, we consider system sizes of: 3 qubits ($\mu + 2$ nn F's), 11 qubits ($+8$ n-nn F's), 21 qubits ($+10$ nn-nn F's), and 29 qubits ($+8$ nnn-nn F's). \\

We utilised a range of simulation techniques, in order to investigate a number of different properties of our proposed algorithm. Simulations of the random-phase-approximation based approach described in Sec.~\ref{Sec:MuonPolFunctions} were carried out using the QuEST package for emulating quantum circuits~\cite{Jones2019quest}. These simulations manually generated the initial state shown in Eq.~(\ref{Eq:Celio_2}), and then carried out quantum circuit emulations of first and second-order Trotter decompositions of the time evolution operator. QuEST is implemented in the C language, and can be efficiently parallelised using OpenMP or MPI. This efficiency enabled us to perform calculations on system sizes of up to 29 qubits, surpassing the largest calculations performed previously in the $\mu^+$SR literature~\cite{huang2012MuSupercLargeSim} (to the best of our knowledge). Because these simulations used the random-phase-approximation approach, sampling noise is present in the results. We also performed quantum circuit-level simulations of running the algorithm on a quantum processor, using the density matrix simulator included in Cirq~\cite{google2020cirq}, a Python package for the simulation of NISQ hardware. Due to the increased computational cost of storing and manipulating the density matrix, these calculations were restricted to systems of up to 11 qubits. However, these simulations enable us to initialise the environment in a maximally mixed state, eliminating the sampling error present in wavefunction-based approches. They also provide a more efficient way to investigate the effects of circuit-level noise on the algorithm. We also performed exact diagonalisation of the Hamiltonians of systems with up to 11 qubits. This provided exact results which can be used to quantify the error introduced by Trotterizing the time evolution operator. The Hamiltonians in this work were generated using OpenFermion, a Python package for generating qubit-mapped Hamiltonians of physical systems, such as molecular electronic structure Hamiltonians~\cite{mcclean2017openfermion}.

\subsection{Noiseless simulations}\label{Subsec:NoiselessResults}

As discussed in Sec.~\ref{Sec:MuonPolFunctions}, there are two sources of algorithmic error in the wavefunction-based method introduced by \textcite{celio1986trotter} for simulating muon polarisation functions; error introduced by Trotterizing the time evolution operator, and sampling errors arising from working with a random-phase augmented wavefunction, rather than with an actual mixed state. In this section, we discuss numerical simulations that quantify the magnitude and scaling of these sources of error, for the CaF$_2 + \mu^+$ system investigated in this work. Given the similarities between Celio's method and the quantum algorithm introduced in Sec.~\ref{Sec:QC_for_muons}, these simulations enable us to better quantify the quantum resources required to run our algorithm. 

\begin{figure}[!h]\centering
\includegraphics[width=0.8\columnwidth]{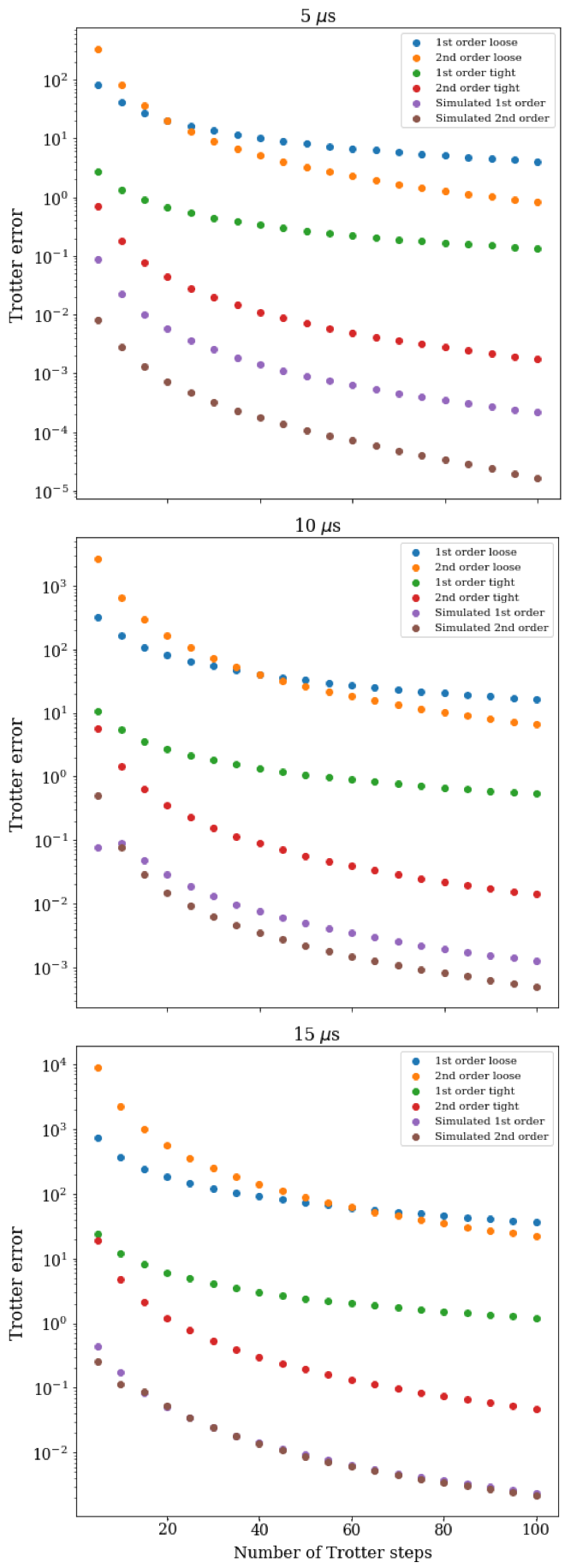}
 \caption{The Trotter error in the three qubit F-$\mu^+$-F system. The error bounds used are described in the main text. In all of these results, we used a magnitude ordering of Hamiltonian terms in the product formula; the terms with the largest coefficients were placed first in the Trotter formula.} \label{Fig:3q_trotter_error}
\end{figure}

\begin{figure}[!h]\centering
\includegraphics[width=0.8\columnwidth]{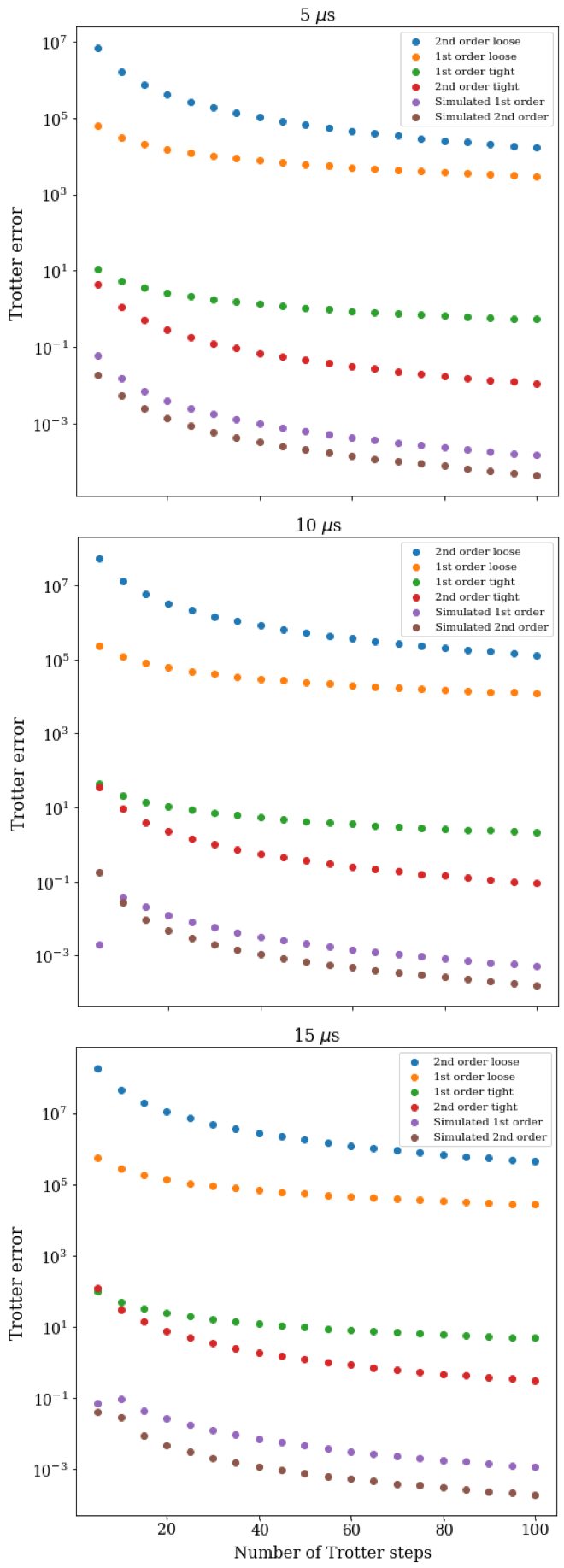}
 \caption{The Trotter error in the 11 qubit system. The error bounds used are described in the main text. In all of these results, we used a magnitude ordering of Hamiltonian terms in the product formula; the terms with the largest coefficients were placed first in the Trotter formula.} \label{Fig:11q_trotter_error}
\end{figure}

\subsubsection{Quantifying errors}
We first investigate the Trotter error present in the algorithm for the 3 and 11 qubit systems. We are able to isolate the Trotter error from the sampling error by simulating the Trotterized time evolution of the density matrix $\frac{I_e}{D_e}\otimes \ket{0}\bra{0}_\mu$. We time evolve this state using both first and second-order Trotter formulae, with a `magnitude ordering' of the terms. By this, we mean that the strongest Hamiltonian terms are placed first in the product formula sequence. We then calculate the polarisation value of the muon qubit at a given time, and compare this to the value obtained from an exact diagonalisation of the muon-environment Hamiltonian (see Eq.~(\ref{Eq:ExactDiag_2})). We compare the numerically simulated Trotter error to two different bounds for the Trotter error. The first bound is given by Eq.~(\ref{Eq:FirstOrderTrotterLoose}) for a first order Trotterization, and Eq.~(\ref{Eq:SecondOrderTrotterLoose}) for second-order Trotter, and represents a loose bound on the error between the exact time evolution operator and the unitary specified by the product formula. The second bound is given by Eq.~(\ref{Eq:FirstOrderTrotterTight}) for a first order Trotterization, and Eq.~(\ref{Eq:SecondOrderTrotterTight}) for second-order Trotter (but with the norm moved inside of the sums in both cases, which loosens the bound), and represents a tighter bound on the error between the exact time evolution operator and the unitary specified by the product formula. The improved `tightness' of the second bound stems from it taking into account commutativity between different terms in the Hamiltonian. These quantities upper bound the error in any observable measured after time evolution, and so will be strictly larger than the error in the polarisation value obtained from our numerical simulations. The Trotter errors for the 3 and 11 qubit systems are shown in Fig.~\ref{Fig:3q_trotter_error} and Fig.~\ref{Fig:11q_trotter_error}, respectively. We plot the Trotter errors obtained at time values of $5, 10,$ and \SI{15}{\micro\second}.

We see from both plots that in all cases the `loose' error bounds are orders of magnitude larger than the `tight' error bounds, which in turn are at least an order of magnitude larger than the numerically observed Trotter errors. As discussed above, this can be partially explained by the fact that the analytic bounds give errors in the unitary evolution, while the numerical results give the error in a specific observable. In addition, despite the improved bounds given by the `tight' formulae, they are still known to not be completely tight to numerical results~\cite{childs2019trottererror}. However, the fact that the tight bounds are still at least an order of magnitude larger than the numerical results highlights the value in work to bound the error in specific observables, rather than existing worst-case bounds. While initial steps have been taken in this direction~\cite{childs2019trottererror, heyl2019quantum}, it would be interesting to consider if tighter bounds are possible for the case of muon spectroscopy, given the simple form of the observable measured. 

Another interesting observation is that both the first and second-order product formulae appear to give similar asymptotic scaling for the numerically observed Trotter error. This is in contrast to the expected behaviour that is evident in the analytic bounds; that the second-order formula should show improved asymptotic scaling as a function of the number of Trotter steps taken. As implementing the second-order formula on a quantum computer can require twice the number of quantum gates needed for the first order formula, this suggests that there may be scenarios in which it is preferable to use the first order approach. For example, the 3 qubit data at \SI{15}{\micro\second} would suggest using a first order decomposition, over a second-order approach. In most cases however, the second-order formula appears to offer improved constant factor scaling that makes up for the increased gate depth required for implementation. It would be interesting to investigate whether it is possible to obtain further improvements using higher-order product formulae, as has previously been observed in the simulation of other spin-$\frac{1}{2}$ systems~\cite{childs2017toward}. We can see that for the 11 qubit system, approximately 40 second-order Trotter steps are sufficient to obtain an accuracy of $10^{-3}$ in the polarisation function at times less than \SI{15}{\micro\second}. Interestingly, the Trotter error for the 11 qubit system does not seem to worsen significantly as the simulated time is increased. \\

\begin{figure}[!h]\centering
\includegraphics[width=0.8\columnwidth]{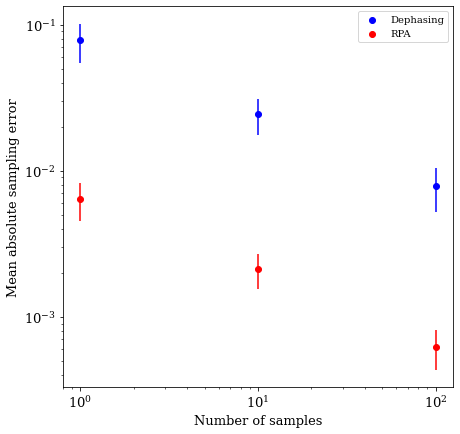}
 \caption{The sampling error present in the random-phase-approximation (RPA) and dephasing methods described in the main text, for the 11 qubit system with 40 second-order Trotter steps. Each point is the average of 20 values taken in the interval $[0, 10)$\SI{}{\micro\second}.} \label{Fig:11q_samplingerror}
\end{figure}

We have also investigated the error present in the random-phase-approximation method that arises from sampling wavefunctions, rather than time evolving the exact mixed state that describes the system. We consider two possible sampling schemes given by Eq.~(\ref{Eq:Celio_2}), which we restate here
\begin{equation}\label{Eq:Celio_2_restated}
\begin{aligned}
\ket{\psi_s(t)} &= U(t) \bigg{[} \frac{1}{\sqrt{D_e}} \sum_j e^{i\theta_j} \ket{0}_\mu \ket{j}_e \bigg{]},
\end{aligned}
\end{equation}
The first scheme, referred to as the `random-phase-approximation (RPA) method' is exactly the same as Celio's method, and considers $\theta_j$ chosen uniformly at random in the range $[0, 2\pi)$. The second, referred to as the `dephasing method' considers $\theta_j$ chosen at random from the discrete set $\{0, \pi\}$. We refer to this approach as the dephasing method, as it can be understood from a quantum computing perspective as first applying a Hadamard gate to each environment qubit to obtain the state
\begin{equation}\label{Eq:EqualSuperposition}
\begin{aligned}
\ket{\psi_s} &= \bigg{[} \frac{1}{\sqrt{D_e}} \sum_j \ket{0}_\mu \ket{j}_e \bigg{]},
\end{aligned}
\end{equation}
and then passing this state through an $N_e$-qubit dephasing channel, with an equal probability of all errors
\begin{equation}\label{Eq:DephasingChannel}
\begin{aligned}
\rho \rightarrow \rho' = \frac{1}{2^{N_e}} \bigg{[} \rho + \sum_{i=0}^{N_e - 1} Z_i \rho Z_i + ... + \prod_{j=0}^{N_e-1} Z_j \rho Z_j \bigg{]}.
\end{aligned}
\end{equation}
This channel can easily be sampled from on a quantum computer by applying any of the possible $Z$-strings acting on $N_e$-qubits with equal probability. This approach can be generalised to particles with spin greater than $\frac{1}{2}$ using the method discussed in Sec.~\ref{Subsec:PrepInitialState}.

In order to isolate the sampling error, we fix the number of Trotter steps, and compare the results obtained from a calculation on the 11 qubit system using both sampling and Trotterization, to those from a calculation that uses Trotterized time evolution of the mixed initial state $\frac{I_e}{D_e}\otimes \ket{0}\bra{0}_\mu$. Without loss of generality, we fix the number of Trotter steps to 40, and take the average of the absolute values of the sampling error at 20 time values, spaced equally in the interval $[0, 10)$\SI{}{\micro\second}. In Fig.~\ref{Fig:11q_samplingerror} we observe that the sampling error arising from the RPA method is around an order of magnitude smaller than that of the dephasing method, due to the increased randomization of the former technique.

\begin{figure}[!h]\centering
\includegraphics[width=0.8\columnwidth]{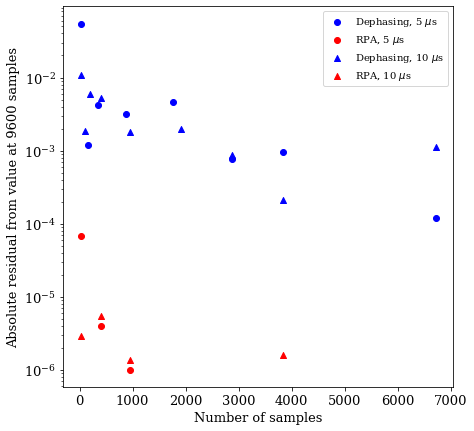}
 \caption{We compare the sampling error obtained using both the random-phase-approximation (RPA) method and the dephasing method, at a given number of samples, to that obtained at 9600 samples. We see that the RPA method rapidly converges to the value obtained at 9600 samples. While we are unable to bound the sampling error at 9600 samples, we see from Fig.~\ref{Fig:11q_samplingerror} that the RPA method can obtain an accuracy of less than $10^{-3}$ for the 11 qubit system with 100 samples. Given that the sampling error decreases with both system size, and the number of samples, we can infer that the error is likely small for even a modest number of samples with the RPA method for the 21 qubit system.} \label{Fig:21q_samplingerror}
\end{figure}

As shown in Eq.~(\ref{Eq:Celio_3}), the error in the sampling based methods decreases as the size of the simulated system increases. We investigated this by considering the sampling error in a larger, 21 qubit simulation. Due to the increased size of this simulation, we were unable to carry out simulations utilising the `exact' mixed initial state for comparison. In lieu of this, we investigate the convergence of the obtained polarisation function as the number of samples is increased. In Fig.~\ref{Fig:21q_samplingerror} we investigate this behaviour for both the RPA and dephasing methods. We observe that the RPA method rapidly converges, and exhibits small fluctuations around its converged value. In particular, we find that using the RPA method, only 20 samples suffice to converge the polarisation function to within $10^{-4}$ of the value obtained with 9600 samples. \\

\begin{figure}[!h]\centering
\includegraphics[width=0.8\columnwidth]{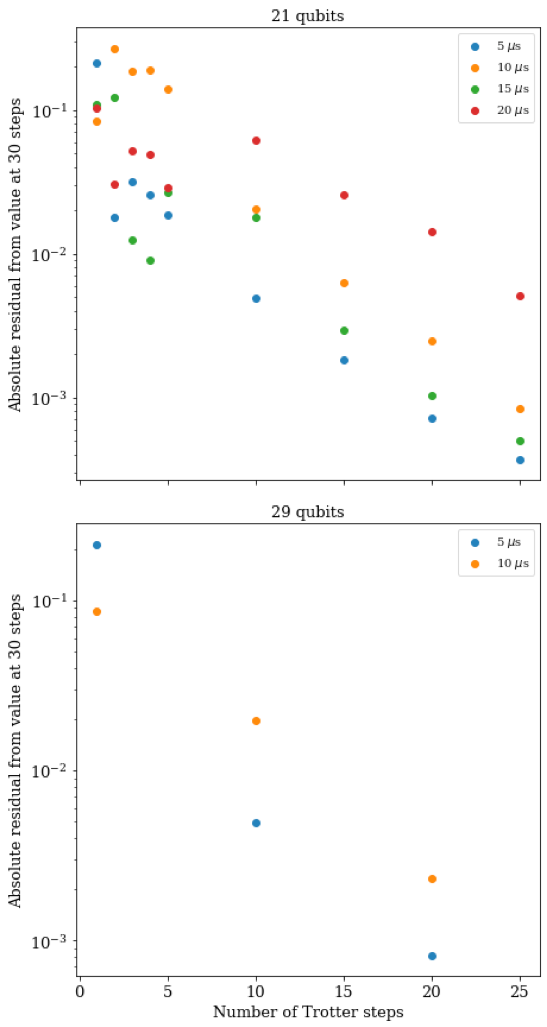}
 \caption{Convergence of the Trotter error for both the 21 and 29 qubit systems, as a function of the number of Trotter steps used. As these systems are too large to exactly simulate classically, we are unable to calculate the Trotter error in the polarisation function. As a result, we plot how the polarisation converges towards the value at 30 Trotter steps (which may still be far from the true value).} \label{Fig:trotter_convergence}
\end{figure}

In a similar vein, we can investigate the convergence of Trotter errors for systems too large to be exactly simulated. In Fig.~\ref{Fig:trotter_convergence} we show how the polarisation function converges as the number of Trotter steps is increased, for both the 21 and 29 qubit systems. These simulations were performed using the RPA method, with 48 samples for the 21 qubit simulation, and a single sample for the 29 qubit simulation. We compare the polarisation function at a range of Trotter steps to that obtained with 30 Trotter steps. We note that this metric does not provide a bound on the Trotter error at 30 Trotter steps. However, given that the convergence error in both cases is monotonically decreasing, and less than $10^{-2}$ by 20 Trotter steps, this may be taken as an indication that the polarisation function is rapidly converging as the number of Trotter steps is increased. \\

\subsubsection{Analysing CaF$_2 + \mu^+$ spectra}
\begin{figure*}
\includegraphics[width=1.8\columnwidth]{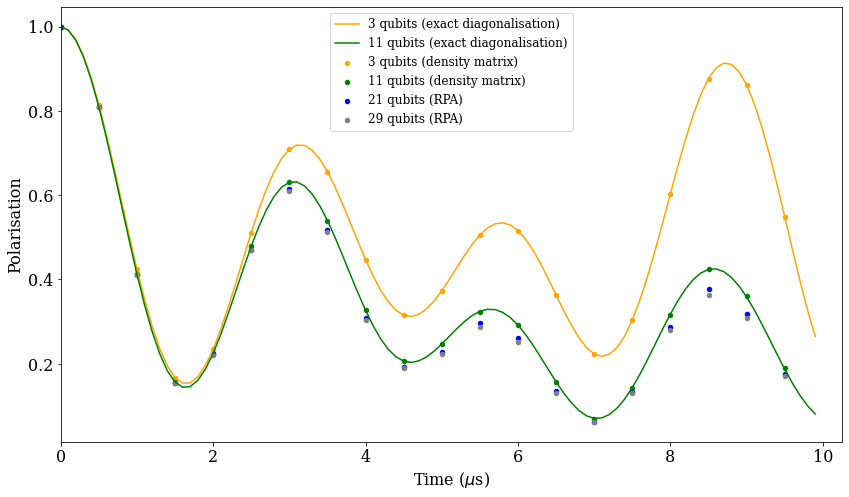}
 \caption{Polarisation functions for the unperturbed 3, 11, 21, and 29 qubit CaF$_2 + \mu^+$ systems. All simulations used 30 second-order Trotter steps to simulate time evolution. The 3 and 11 qubit results were obtained using a density matrix simulation, that eliminates sampling error. The 21 and 29 qubit simulations were performed using the random-phase-approximation method, with 48 and 1 samples, respectively.} \label{Fig:PolarisationSimulation}
\end{figure*}

We can use the results discussed above to determine how many samples and Trotter steps to use in order to quantitatively investigate the CaF$_2 + \mu^+$ system. In Fig.~\ref{Fig:PolarisationSimulation} we plot the angular averaged polarisation functions obtained for the 3, 11, 21, and 29 qubit CaF$_2 + \mu^+$ systems for the first \SI{9.5}{\micro\second} of evolution. The simulated data points for the 3 and 11 qubit systems were obtained by time evolving the initial state $\frac{I_e}{D_e}\otimes \ket{0}\bra{0}_\mu$, using 30 second-order Trotter steps, which suffices to measure the polarisation function to an accuracy of less than $10^{-2}$. These data points are plotted with the polarisation functions obtained from exact diagonalisation of the corresponding system Hamiltonian. The simulated data points for the 21 and 29 qubit systems were obtained using the RPA method, with 48 samples used for the 21 qubit system, and a single sample used for the 29 qubit system. Both of these simulations also used 30 second-order Trotter steps. We see from Fig.~\ref{Fig:PolarisationSimulation} that introducing additional fluorine nuclei causes a damping effect on the polarisation function, which appears well converged at 29 qubits. This observation is in good agreement with the recent numerical results of \textcite{wilkinson2020muonfluorinefull}, who showed that the additional environmental spins act as a source of decoherence, as polarisation leaks from the muon to the environment. \\

Having confirmed that our algorithm produces the qualitative results expected, we can use it to locate the muon rest site in CaF$_2$. This is achieved by parametrizing the muon-fluorine distances, and generating a polarisation function for a given geometry. We can then compute the value of a loss function between the simulated data and the experimental data (in this case, the reduced-$\chi^2$ value), and use an optimisation algorithm to generate new values of the parameters that describe the geometry of the system. In our numerical simulations, we used the Nelder-Mead algorithm. It was necessary to use a derivative free optimisation method because of the sampling noise present in the RPA method, which is larger than the finite-difference steps used to calculate the gradient in many black-box optimisation algorithms. We used this approach to optimise the geometry of the 21 qubit system (composed of the muon, the two nearest-neighbour fluorines, 8 next-nearest-neighbour, and 10 next-next-nearest-neighbour fluorines). Five geometric fitting parameters were used, which we describe via the colours used in Fig.~\ref{Fig:System_Geometry}d:
\begin{enumerate}
    \item The distance between the muon (red) and the nearest-neighbour fluorines (both coloured black).
    \item The distance between the muon and the next-nearest-neighbour fluorines (all coloured blue).
    \item The distance between the muon and the green next-next-nearest-neighbour fluorines.
    \item The distance between the muon and the purple next-next-nearest-neighbour fluorines.
    \item The distance between the muon and the orange next-next-nearest-neighbour fluorines.
\end{enumerate}
along with two parameters to fit the asymmetry
\begin{equation}\label{Eq:FittingAsym}
    A(t) = A_0 P(\theta_1, ... , \theta_5; t) + A_{\mathrm{bg}} 
\end{equation}
We used the RPA method to generate the polarisation function, with 40 second-order Trotter steps, and a single sample for each data point.

We plot the fit to the experimental data in Fig.~\ref{Fig:fitting_exp}. The fitted data shows excellent qualitative agreement with the experimental data, particularly at early times. The fit obtains a reduced $\chi^2$ value of 2.13, which suggests an incomplete fit to the data. We attribute this to a combination of limited experimental data at times beyond \SI{10}{\micro\second}, as well as the ineffective nature of the Nelder-Mead optimisation algorithm, which is liable to becoming trapped in local minima. Our fitting procedure causes the nearest-neighbour fluorines to move towards the muon by $0.188$\AA. This is in excellent agreement with the calculation of \textcite{wilkinson2020muonfluorinefull}, which yielded a value of $0.190(1)$\AA~for the same quantity. We note that the results of Ref.~\cite{wilkinson2020muonfluorinefull} were obtained by fitting an 11 spin system to the experimental data, and considering two physical parameters; the muon -- nearest-neighbour fluorines distance, and a factor that scaled the strength of the next-nearest-neighbour interactions to act as a proxy for more distant nuclei. Our fitting procedure also caused the next-nearest-neighbour fluorines to move towards the muon by $0.206$\AA. While Ref.~\cite{wilkinson2020muonfluorinefull} did not explicitly consider the effect of perturbing the positions of the more distant nuclei, they carried out density functional theory (DFT) calculations suggesting that the next-nearest-neighbour fluorines only move towards the muon by around $0.03$\AA. We carried out additional numerical simulations with this geometry for the 21 spin system (shown in Fig.~\ref{Fig:fitting_dft} in Appendix~\ref{Appendix:AdditionalResults}), and found a worse fit for the experimental data, with a $\chi_\nu^2$ value of 4.44. This implies that either:

\begin{figure*}
\includegraphics[width=1.8\columnwidth]{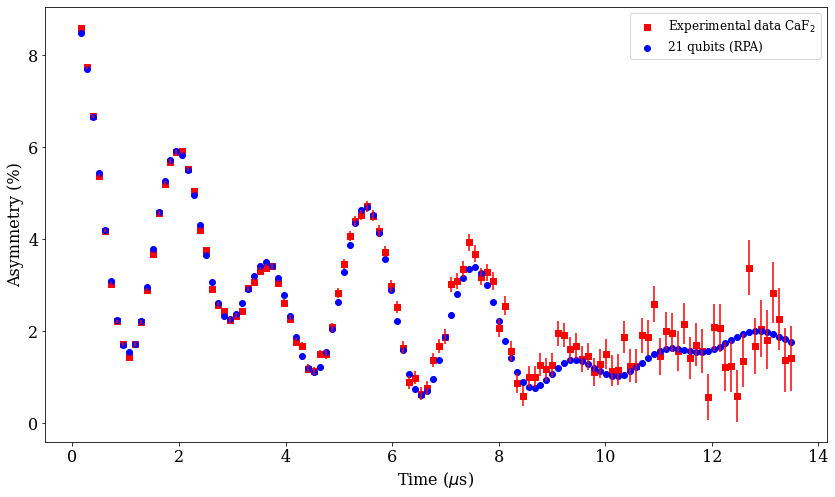}
 \caption{A fit of the 21 qubit simulated data (using the random-phase-approximation method with 1 sample per data point, and 40 second-order Trotter steps) to the experimental data obtained in Ref.~\cite{wilkinson2020muonfluorinefull}. The fit was performed using the gradient-free Nelder-Mead algorithm. The fitting parameters were the muon-fluorine distances described in the main text, and a scaling factor and offset to convert the simulated polarisation value to an asymmetry value.} \label{Fig:fitting_exp}
\end{figure*}

\begin{enumerate}
    \item Our system size of 21 spins is still not large enough to fully capture the continuum extrapolation of Ref.~\cite{wilkinson2020muonfluorinefull}, and larger simulations are required (which are impractical to perform on classical hardware).
    \item The fit obtained by our simulation may suggest an inaccuracy in the DFT results. However, from a physical perspective, it would be surprising if the more distant nuclei were more strongly attracted towards the muon than the nearest-neighbours.
\end{enumerate}
The most likely explanation may be a combination of these factors, together with the fact that a number of similarly good local minima are present in the $\chi^2$ surface for our parameter space. This highlights the value in utilising complementary techniques to analyse muon spectroscopy data, as well as the benefit provided by having access to as large a simulation of the system as is possible.

\subsection{Noisy simulations}\label{Subsec:NoisyResults}

\begin{figure}[!h]
\includegraphics[width=0.8\columnwidth]{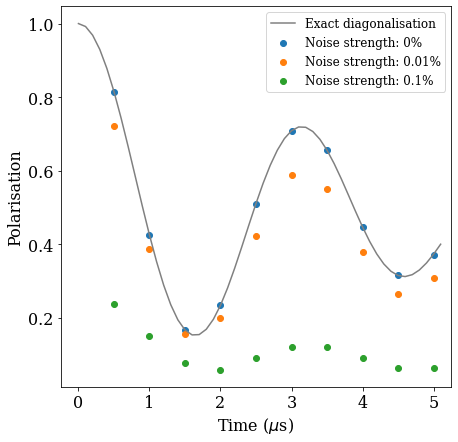}
 \caption{Density matrix simulations of the three qubit F-$\mu^+$-F system, with single qubit depolarising noise applied after each gate. The noise channel is applied independently to each qubit involved in the gate. The simulation used 20 second-order Trotter steps, which yielded a circuit with 900 single-qubit gates, and 680 two-qubit gates.} \label{Fig:NoisySimData}
\end{figure}

As discussed in Sec.~\ref{Sec:QuantumComputingIntro}, existing quantum computers have far higher error rates than their classical counterparts. The presence of physical noise can corrupt the results of calculations on these devices, rendering any quantum speedup offered moot. As such, it is essential to investigate the effects of noise on our algorithm. A simple model of the noise present in near-term quantum computing devices is the single qubit depolarizing noise channel
\begin{equation}\label{Eq:DepolNoise}
    \rho \rightarrow \rho' = (1-p)\rho + \frac{p}{3}(X \rho X + Y \rho Y + Z\rho Z)
\end{equation}
where $\rho$ and $\rho'$ are the density matrix of the qubit before and after the noise channel (respectively). This channel effectively applies an $X, Y$ or $Z$ error to the qubit, with probability $p/3$, and leaves it unchanged with probability $(1-p)$. We assume that this noise model is applied after each gate in the circuit, acting separately on each qubit involved in the gate. In Fig.~\ref{Fig:NoisySimData} we plot the effects of this depolarizing noise model on an exact density matrix simulation of the 3 qubit F-$\mu^+$-F system. We use 20 second-order Trotter steps for the simulation, which suffices to obtain an accuracy of less than $10^{-3}$ in the polarisation function at all values within the first \SI{5}{\micro\second}. We merge adjacent single qubit gates together, to yield a circuit with 900 single qubit gates, and 680 two-qubit gates. In this error model, we are assuming that noise is uncorrelated between the qubits involved in a two-qubit gate, and that the error rate for single-qubit gates is the same as that for two-qubit gates. We note that the circuit depth is the same for all of the points calculated. This means that the Trotter error is likely smaller in points taken at earlier times than at later times, while the effective physical noise rate of the circuit will be similar at all points. In reality, we would likely choose to fix the Trotter error along the polarisation function, which would enable us to use a shorter depth circuit to simulate points at earlier times -- thus reducing the physical noise in those datapoints. We see from Fig.~\ref{Fig:NoisySimData} that even with a depolarizing noise rate of $p=10^{-4}$ (which is an order of magnitude lower than the best error rates observed to date in quantum hardware~\cite{gaebler2016high, ballance2014high}), there is still a significant decay in the value of the simulated polarisation function. This noise rate corresponds to an expected number of errors per circuit (defined as the error rate multiplied by the number of gates in the circuit) of around 0.16.

\begin{figure}[!h]
\includegraphics[width=0.8\columnwidth]{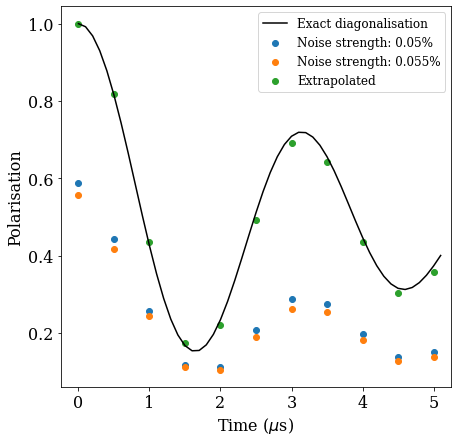}
 \caption{Using exponential extrapolation of data obtained at two different noise rates to infer the noise free polarisation function. The extrapolation was performed using Eq.~(\ref{Eq:ExponentialExtrap}). The data points were obtained in the same way as described in the caption of Fig.~\ref{Fig:NoisySimData}. The expected number of errors in the circuit was 0.79.} \label{Fig:ExtrapResults}
\end{figure}

It has been previously observed that when the expected number of errors in the circuit is less than around unity, the error mitigation techniques introduced in Sec.~\ref{Sec:QuantumComputingIntro} can be successful in recovering the noiseless value of the circuit~\cite{endo2017practical, cai2020extrapolation}. In this work, we investigate the use of exponential extrapolation to mitigate the effects of noise. We consider the same 3 qubit system discussed above, with a baseline noise rate of $p=5 \times 10^{-4}$. We then artificially `boost' the error rate by a heuristically chosen factor of $\lambda = 1.1$, and calculate the extrapolated expectation value as
\begin{equation}\label{Eq:ExponentialExtrap}
    P_0 = \bigg{(}\frac{P_\epsilon^\lambda}{P_{\lambda \epsilon}}\bigg{)}^{\frac{1}{\lambda - 1}},
\end{equation}
where $P_\epsilon$ is the polarisation value calculated with the baseline noise rate, and $P_{\lambda \epsilon}$ is the polarisation value calculated with the boosted noise rate. We see from Fig.~\ref{Fig:ExtrapResults} that this exponential extrapolation is able to recover almost noiseless results, despite the large damping of the polarisation function in the unmitigated case. The noise strength of $p=5 \times 10^{-4}$ corresponds to an expected number of errors per circuit of 0.79. The mean absolute error in polarisation function after exponential extrapolation is 0.011. \\ 

In this work, we are investigating a quantum algorithm for analysing muon spectroscopy data. As such, it is interesting to ask whether the noise inevitably present in $\mu^+$SR experimental results means that physical errors in the quantum computer are less damaging than they would be in other applications (such as trying to calculate the ground state energies of chemical systems). As shown in Fig.~\ref{Fig:fitting_exp}, muon spectroscopy data can exhibit errors arising from a number of sources - most notably, the statistical uncertainties arising from recording a finite number of muon decay events. In order to investigate the noise robustness of our algorithm for analysing $\mu^+$SR data, we try to locate the muon rest site in the presence of noise in the simulated data for the 11 qubit system. Rather than considering the circuit-level noise models discussed above, we effectively engineer noise in the polarisation function through undersampling in the RPA method. As shown in Fig.~\ref{Fig:11q_samplingerror}, we can generate mean errors of 0.0243, 0.0068, and 0.0022 in the polarisation function by using the dephasing method with 10 samples, the RPA method with 1 sample, and the RPA method with 10 samples, respectively. We note that there are two important assumptions present in this noise model. Firstly, we are assuming that physical noise would cause the data points to become normally distributed around the noiseless value, as happens for the sampling noise. As is evident from Fig.~\ref{Fig:NoisySimData}, this is not necessarily the case; many sources of noise will simply cause a decay in the polarisation function. However, we can see from Fig.~\ref{Fig:ExtrapResults} that performing extrapolation can change how the data points are distributed around the noiseless result. When shot noise in the quantum algorithm is taken into account, this may enable us to approach the normal distribution of noise considered in these simulations. Another path to recover normally distributed results could be to use an alternative error mitigation method, known as the `quasiprobability technique'~\cite{temme2017mitigation, endo2017practical, huo2018temporally, strikis2020quasiproblearning}. The quasiprobability technique attempts to invert the physical noise channel of each gate, by sampling from an increased number of carefully modified, noisy circuits, and combining the results in post-processing. This approach has previously been observed to give results that are approximately normally distributed around the true values~\cite{endo2017practical}. Secondly, we are assuming that the magnitude of the noise is the same for all data points. In reality, as discussed above, we would fix the Trotter error along the polarisation function, and thus consider shorter depth circuits at earlier times. This would reduce the noise in earlier data points, mimicking the error bars present in the experimental results (in the experimental case, the muon half life is around \SI{2.2}{\micro\second}, so a significant number of decay events must be observed to record sufficient statistics at later time values). In our simulations we use 40 second-order Trotter steps, and consider two geometric fitting parameters:
\begin{enumerate}
    \item The muon - nearest-neighbour fluorines distance.
    \item The muon - next-nearest-neighbour fluorines distance.
\end{enumerate}
along with two parameters to fit the asymmetry
\begin{equation}\label{Eq:FittingAsym2}
    A(t) = A_0 P(\theta_1, \theta_2; t) + A_{\mathrm{bg}} 
\end{equation}
Once again, we use the Nelder-Mead algorithm to minimise the error-weighted least-squares residuals between the experimental data and the simulated data. 

\begin{figure}[!h]
\includegraphics[width=0.8\columnwidth]{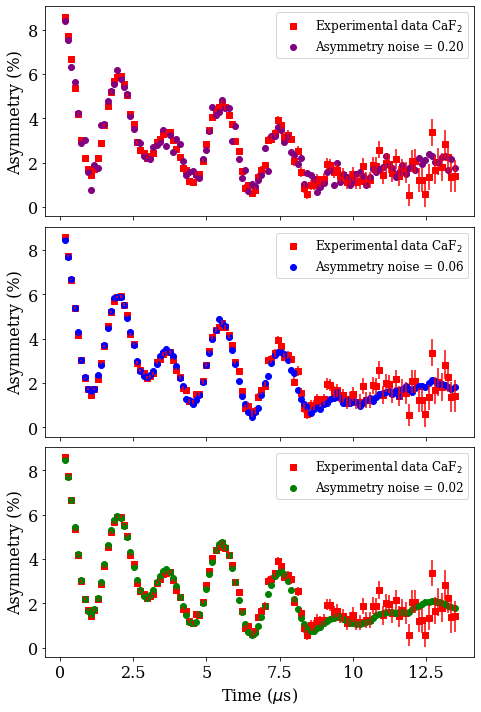}
 \caption{Fits of the 11 qubit simulated data with 40 second-order Trotter steps to the experimental data obtained in Ref.~\cite{wilkinson2020muonfluorinefull}. The noise in each simulated datapoint was varied by controlling the number of samples used. The fit was performed using the gradient-free Nelder-Mead algorithm. The fitting parameters were the muon-fluorine distances described in the main text, and a scaling factor and offset to convert the simulated polarisation value to an asymmetry value.} \label{Fig:NoisyAsymmetry}
\end{figure}

In Fig.~\ref{Fig:NoisyAsymmetry} we plot examples of the fitted data obtained at the three noise values listed above. We see that even at high noise rates of around 2.5\% in each data point, the fitting procedure is able to approximately capture the overall shape of the polarisation function. As the strength of the noise is reduced, we obtain increasingly better fits to the experimental data.

\begin{figure}[!h]
\includegraphics[width=0.7\columnwidth]{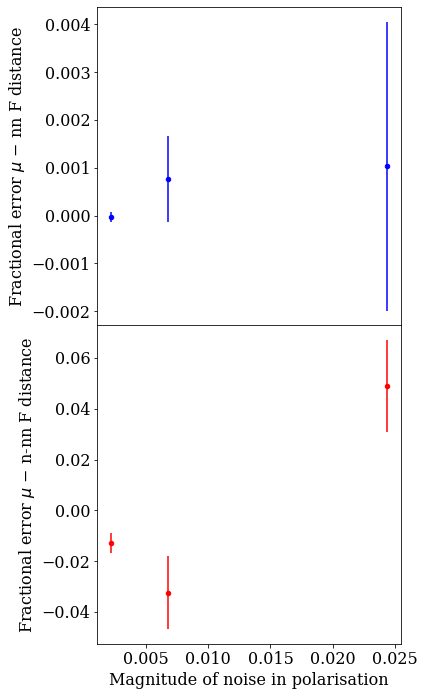}
 \caption{The fractional error in the geometric fitting parameters used to fit the simulated muon spectroscopy data to the experimental data in Fig.~\ref{Fig:NoisyAsymmetry}. The noiseless values of the parameters were obtained through exact diagonalisation of the 11 qubit Hamiltonian to obtain a polarisation function, which was fitted using the Nelder-Mead algorithm.} \label{Fig:NoisyParameters}
\end{figure}

In order to better quantify the degree of noise robustness present in the algorithm, we plot the error that noise induces on the noiseless values of the two geometric fitting parameters in Fig.~\ref{Fig:NoisyParameters}. The noiseless values of the fitted parameters were obtained by fitting the 11 qubit polarisation function, generated by exact diagonalisation, to the experimental data, using the Nelder-Mead algorithm. The upper plot in Fig.~\ref{Fig:NoisyParameters} shows that as the average noise in each individual data point is reduced, we obtain an improved estimate of the noiseless muon -- nearest-neighbour fluorine distance. Interestingly, the fractional error in the parameter is around an order of magnitude smaller than minimum fractional error in the polarisation value (the maximum value of the polarisation function is 1, so the minimum fractional error is equivalent to the noise strength of the simulation). We attribute this noise resilience to the fact that the fitting procedure is extracting a global property of the system (the muon -- fluorine distance) rather than the value of any individual data point. The lower plot in Fig.~\ref{Fig:NoisyParameters} suggests that the muon -- next-nearest-neighbour fluorine distance is less resilient to noise, as the fractional error in the parameter is approximately the same order of magnitude as (but larger than) the magnitude of the noise in the polarisation function. Nevertheless, we are optimimistic that the algorithm could be made more noise resilient by: 1) Reducing the noise strength at earlier times by adapting the number of Trotter steps used (as discussed above), and 2) Using an optimisation routine known to be more resilient to noise than the Nelder-Mead algorithm, such as Bayesian optimisation.

Linking back to our earlier simulations of circuit-level noise, we note the following:
\begin{enumerate}
    \item Our fitting routine obtained accurate parameter values when the error in each simulated polarisation data point was less than around 0.01 (for the 11 qubit system).
    \item Using exponential extrapolation, we obtained a mean absolute error in the polarisation function of 0.011 when the expected number of errors in the circuit was around 0.8 (for the 3 qubit system).
\end{enumerate}
Assuming that the circuit-level result holds for larger system sizes (we would expect the results to be better for larger system sizes, as when errors do occur, in larger system sizes they are less likely to occur on the crucial muon qubit), we infer that our data analysis algorithm is able to tolerate an expected error rate of around 0.8 errors per circuit, on average. In the 11 qubit case, this would enable us to learn the muon -- nearest-neighbour fluorine distance to an accuracy better than 0.1\%, and the muon -- next-nearest-neighbour fluorine distance to an accuracy of around 5\%. 

It would be interesting to consider whether this fitting procedure could be shown to be provably noise robust. If so, we would expect these results to be useful beyond this work, earmarking experimental data analysis as a promising application area for quantum computers, even in the presence of noise.

\subsection{Resource estimates}\label{Subsec:ResourceEstimates}

We now consider the quantum resources required to perform the calculations described above. We consider different cost metrics for NISQ simulations and error corrected simulations. When performing NISQ calculations, we aim to minimise both the circuit depth and the total number of gates, as these will determine the number of errors that are likely to occur in the circuit. In particular, we focus on two-qubit gates, as in many systems these have error rates at least an order of magnitude worse than those of the single qubit gates. As NISQ devices typically lack the resources required for quantum error correction, the number of gates that can be applied in the circuit is limited by the magnitude of the noise in the system.

In order to determine an appropriate fault tolerant cost metric, we first need to consider which error correcting code we are considering using for the simulation. One possible choice of error correcting code is the surface code~\cite{kitaev1997quantum}. The surface code has received considerable attention in recent years, owing to its comparatively high code-threshold of around 1\%, and its compatibility with 2D nearest-neighbour connectivity architectures~\cite{fowler2012surface} -- which are realistic architectures for solid state qubit systems. When considering error corrected calculations using the surface code, we seek to minimise the number of T gates. While T gates (or an alternative `non-Clifford group' gate) are required for universality in quantum computation, these gates cannot be natively implemented in a fault-tolerant manner in the surface code. They are typically implemented via processes known as `magic state distillation' and `magic state injection', and are often the dominant cost in fault-tolerant resource estimates. Arbitrary angle single-qubit rotations can be synthesised from these T gates. The number of T gates required per rotation depends logarithmically on the inverse of the synthesis error~\cite{ross2016CliffordTcompilations}. A reasonable assumption for calculations of the size considered in this work, is that around $100$ T gates per single-qubit rotation will suffice~\cite{pallister2020JwGadget}.

\subsubsection{NISQ resource estimates}\label{Subsubsec:NISQ_resources}
As shown in Fig.~\ref{Fig:11q_trotter_error}, we require approximately 40 second-order Trotter steps to reduce the Trotter error in the observable to $10^{-3}$ when calculating $P(t=$\SI{15}{\micro\second}) for the 11 qubit system. We perform a basic compilation of the circuit, by merging adjacent single-qubit gates. This results in a gate count of $5\times 10^4$ single-qubit gates, and $3.9 \times 10^4$ two-qubit gates. If we assume that single-qubit gates have an error rate ten times lower than two-qubit gates, then this circuit is roughly equivalent to implementing $4.4 \times 10^4$ two-qubit gates. As discussed in Sec.~\ref{Subsec:NoisyResults}, we observed that for the 3 qubit system, exponential extrapolation was able to recover acceptably accurate results when there were around 0.8 expected errors per circuit. In order to achieve this circuit error rate for the 11 qubit system with $4.4 \times 10^4$ two-qubit gates, we would require a two-qubit gate error rate of $p=2 \times 10^{-5}$. This is approximately two orders of magnitude lower than the current lowest two-qubit gate error rates~\cite{gaebler2016high, google2019supremacy, ballance2014high}. Simulations of larger, classically challenging system sizes will require an even larger number of gates. The most promising avenue for realising these calculations before the advent of quantum error correction is to significantly reduce the gate count required. One possible route towards this goal would be to consider Trotter orderings that reduce the Trotter error, or that enable a larger number of gates to be cancelled. We could alternatively consider neglecting the dipolar interactions between the nuclei, retaining only the muon-fluorine dipolar interactions (as has previously been done in classical simulations~\cite{huang2012MuSupercLargeSim}). Even if these optimisations can be incorporated, it appears challenging for NISQ devices to surpass classical $\mu^+$SR simulation capabilities. This is not necessarily reflective of the performance of the algorithm described in this work, and instead serves to highlight both the quality of classical simulations of quantum systems, and the challenges inherent in many NISQ algorithms, resulting from high noise rates in current quantum processors. \\

\subsubsection{Error corrected resource estimates}\label{Subsubsec:FT_resources}
Given the challenges faced by NISQ machines, it is natural to investigate the resources required to implement the algorithm described in this work in an error corrected setting. As discussed above, we consider error correction within the surface code. We initially consider running our algorithm on a `small' fault-tolerant quantum computer, meaning that we use the smallest footprint possible to implement the algorithm. Nevertheless, this device will likely still contain many thousands of physical qubits. We follow the approach to surface code resource estimation taken in Ref.~\cite{Litinski2019gameofsurfacecodes}, which considers surface code operations implemented using lattice surgery, complemented with magic state distillation for implementing T gates. We use a `compact' data block of physical qubits to store each logical qubit. For a system with $Q$ logical qubits, using a compact block results in needing $\lceil 1.5Q + 3 \rceil$ surface code tiles. For a distance $d$ surface code, each surface code tile consists of $2d^2$ physical qubits ($d^2$ data qubits and $d^2$ syndrome qubits).

For the 11 qubit system with 40 second-order Trotter steps, we require $1.96 \times 10^4$ non-Clifford single-qubit rotations, which we assume requires $1.96 \times 10^6$ T gates. We use a 15-1 distillation block for magic state distillation, which produces output magic states with an error rate of at most $3.5 \times 10^{-8}$, and requires an additional 11 surface code tiles. For our 11 qubit system, this results in $B=31$ surface code tiles in total. With this setup, we can consume a magic state every $11d$ surface code cycles. Our calculation therefore must `survive' for $T = 11d \times (1.96 \times 10^6)$ code cycles. The size of the surface code required to perform our calculation can be determined using~\cite{Litinski2019gameofsurfacecodes}
\begin{equation}\label{Eq:SurfaceCodeEstimates}
    B \cdot T  \cdot 0.1(100p)^{(d+1)/2} < \epsilon 
\end{equation}
where $p$ is the physical error rate, and $\epsilon$ is the target circuit error rate. The fault-tolerant resources required are shown in Table.~\ref{Tab:FT_resources}, for a number of scenarios. We consider resource estimates with a realistic two-qubit gate error rate of $p=10^{-3}$, as well as a more optimistic error rate of $p=10^{-4}$. We also consider two target noise suppression levels. In the first case, we consider suppressing the error rate such that there are $\epsilon=0.01$ errors in the circuit, on average. In the second case, we assume that we are able to tolerate a noise rate of $\epsilon=0.8$ errors in the circuit, on average. This choice of target circuit error rate is motivated by the success of exponential extrapolation in obtaining accurate results for the 3 qubit system in the presence of noise of this magnitude. We assume a surface code decoding cycle time of \SI{1}{\micro\second}. We can see from Table.~\ref{Tab:FT_resources} that while the error robustness of the algorithm can reduce the resources required by a factor of around 1.5, this is overshadowed by the exponential improvements arising if the physical error rate can be reduced.

\begin{center}
\begin{table}
\begin{tabular}{ c|c|c|c|c }
 p & $\epsilon$ & d & Physical qubits & Circuit time (s)\\
 \hline
 $10^{-3}$ & 0.01 & 22 & 30,008 & 474 \\ 
 $10^{-4}$ & 0.01 & 10 & 6200 & 216 \\
 $10^{-3}$ & 0.8 & 18 & 20,088 & 388 \\
 $10^{-4}$ & 0.8 & 8 & 3968 & 172 \\
\end{tabular}
\caption{\label{Tab:FT_resources} Error corrected resource estimates for simulations of the 11 qubit CaF$_2 + \mu^+$ system with 40 second-order Trotter steps. Here, $p$ is the physical error rate, $\epsilon$ is the expected number of errors per circuit, and $d$ is the surface code distance used. }
\end{table}
\end{center}

We can also estimate the resources required for a larger system, such as the 29 qubit system. We assume that 50 second-order Trotter steps are needed, which leads to a circuit involving around $2.3 \times 10^{7}$ T gates. We use the same surface code setup as described above, and consider a physical noise rate of $p=10^{-3}$, and a target circuit error rate of $\epsilon=0.8$. We find that a distance 21 surface code is needed, requiring around 51,000 physical qubits. The calculation takes 5234 seconds to run. On first inspection, this is a modest number of physical qubits, compared to existing resource estimates for solving classically challenging chemistry calculations with quantum hardware. At $p=10^{-3}$ physical error rates, it has previously been estimated that around $3 \times 10^5$ physical qubits would be required to find the ground state energies of classically challenging instances of the Fermi-Hubbard model~\cite{kivlichan2019condensedtrotter}, and that over a million physical qubits running for several days may be needed to find the ground state of small molecules relevant for catalysis~\cite{vonburg2020MicrosoftQubitizedCatalysts, lee2020tensor_hypercontraction_qubitization}. We note that these resource estimates were upper bounds that may be loose (particularly in the case of the Fermi-Hubbard resource estimates, which used loose analytic bounds on the Trotter error), while our estimates used tighter error bounds obtained from numerical simulations. Moreover, the calculations considered in these other works are likely more challenging than the 29 qubit simulations discussed herein. Nevertheless, these mitigating factors are unlikely to fully account for the reduced resources needed for our approach. We attribute some of these resource savings to the fact that we are simulating spin systems, which may require less overhead than fermionic systems when mapping to qubits. Moreover, we are able to exploit the noise robustness exhibited by our algorithm to further reduce the surface code distance required. \\

Unfortunately, there are limitations to the approach taken in our resource estimates above, that cause us to reconsider how these calculations would be performed on error corrected quantum computers. As seen in Table.~\ref{Tab:FT_resources}, performing a single iteration of even the 11 qubit simulation would take several minutes, assuming a surface code decoding cycle time of \SI{1}{\micro\second} (which may be optimistic for slower systems, such as trapped ion quantum computers). If we estimate the polarisation value through direct sampling of the wavefunction, then obtaining a precision of $10^{-2}$ in the polarisation would take around 10,000 repetitions. We must then repeat the calculation at a number of simulated time values to perform a single step of our optimisation subroutine. In turn, the optimisation routine may require hundreds of iterations in order to obtain a good fit for the experimental data. As a result, if the methods discussed in this work are to prove useful for analysing experimental data, we must find ways to reduce the runtime of the algorithm.

As discussed in Sec.~\ref{Subsec:Measurement}, we can obtain a measurement of the polarisation using a combination of amplitude amplification and quantum phase estimation. This approach increases the circuit depth required by a factor of $1/\epsilon_m$ (where $\epsilon_m$ is the measurement precision desired), but reduces the number of repetitions required to a constant value. While a full resource estimate of using this technique is beyond the scope of this work, we can perform a rough estimate of how this approach would compare to those discussed above. We assume a value of $\epsilon_m = 10^{-2}$, increasing the circuit depth required by a factor of 100. For the 29 qubit system, this increases our T count estimate to $2.3 \times 10^{9}$ T gates. We neglect the resources required for magic state distillation, and focus on the resources required to reduce the expected number of errors in the logical qubits to less than $\epsilon = 0.01$. Assuming that distilling a magic state still takes $11d$ surface code cycles, we need to solve
\begin{equation}\label{Eq:SurfaceCodeEstimates29}
    \lceil (1.5 \times 29) + 3 \rceil \cdot (11 \times  2.3 \times 10^{9})  \cdot 0.1(100p)^{(d+1)/2} < 0.01 .
\end{equation}
This expression is satisfied for $d \geq 29$. The compact data block for the 29 logical qubits corresponds to around 98,000 physical qubits. With this minimal setup, the calculation would take at least 8 days to complete. As discussed in Ref.~\cite{Litinski2019gameofsurfacecodes}, we can reduce the runtime of the calculation by exchanging spatial resources for time resources. Adding more qubits enables us to distill and consume magic states more quickly, until we reach the point at which a magic state can be consumed every surface code clock cycle. The addition of further qubits allows us to parallelise the implementation of T gates using quantum teleportation. If the T gates required by our algorithm can be fully parallelised into around $8\times 10^7$ layers, and one layer can be implemented per surface code cycle, a calculation of the polarisation value along a single axis, at a single point in time, would finish in 80 seconds. Building a quantum computer with these capabilities would require billions of qubits~\cite{Litinski2019gameofsurfacecodes} -- considerably larger than any machines that are planned within the coming decades.\\

Given the challenges associated with building even a small error corrected quantum computer, it is natural to ask whether there are other ways to reduce the runtime of the algorithm. First and foremost, we should search for algorithmic improvements. As in the NISQ case, better compilation routines, or the use of more efficient time evolution algorithms, could reduce the number of gates required to execute our proposal. In particular, we could consider alternative Hamiltonian decompositions that reduce the Trotter error by grouping commuting terms~\cite{vandenBerg2020circuitoptimization}, or that exploit the locality of power law interactions~\cite{tran2019powerlaw}. Similarly, we could utilise techniques that exploit symmetries in the Hamiltonian coefficients to reduce the total number of T gates required for the circuit~\cite{Gidney2018halvingcostof,kivlichan2019condensedtrotter}. 

Given the embarrassingly parallel nature of our algorithm, it may have to wait until quantum hardware becomes as cheap and ubiquitous as classical computing is today. This would enable us to distribute each data point and repetition across a number of different processors -- in much the same way as the classical emulations in this work were performed. Finally, we note that the challenges discussed in this section are not unique to our proposal, but will be faced by all quantum algorithms that require a large number of repetitions to fulfil their purpose. Perhaps such `multi-repetition' algorithms will be impractical, unless we are able to develop improved error correcting codes, more reliable physical qubits, or reduced code cycle times -- which would all improve the effective clock speed of the quantum processor.

\section{Conclusion}\label{Sec:Conclude}
In this work, we have introduced and investigated a quantum algorithm for analysing the data arising from muon spectroscopy experiments. As discussed in Sec.~\ref{Sec:MuonSpectro}, muon spectroscopy is a versatile experimental technique, that has been used to investigate a wide range of physical systems and phenomena. In some cases, accurately analysing the spectra produced by these experiments requires comparison to data generated by a quantum model for the system. The cost of simulating these quantum models scales exponentially with the size of the simulated system, using all known classical methods. These techniques are particularly necessary when trying to locate the muon rest site, or when studying the effects of muon diffusion.

The quantum algorithm introduced in this work is quite simple, and resembles the (classical) random-phase-approximation based method introduced to the muon community by \textcite{celio1986trotter}. Our algorithm constructs suitable initial states on the quantum computer, evolves them in time, and measures the $Z$ expectation value of the muon qubit, in order to construct simulated polarisation functions. The number of gates required by our algorithm scales polynomially with the size of the system, at worst as $\mathcal{O}(N^5)$, as shown in Sec.~\ref{Subsec:TimeEvolution}.

Numerical emulations of our quantum algorithm on classical hardware have enabled us to bound the error from both Trotterization, and finite sampling of the initial mixed state. We observed that the error in the polarisation function resulting from Trotterization was orders of magnitude smaller than existing analytic bounds. We applied these numerical emulations to analyse the data from a muon spectroscopy experiment on CaF$_2$, and observed good agreement with the recent analysis of \textcite{wilkinson2020muonfluorinefull} for the same dataset, which used state-of-the-art classical simulation methods. In the process, we performed the largest simulation employed in muon spectroscopy analysis to date, with a Hilbert space size of $2^{29}$. 

By considering the impact of noise in the quantum computer on our algorithm, we have been able to estimate the quantum resources required to perform classically challenging instances of muon spectroscopy analysis. In particular, we observed significant noise robustness in our algorithm, stemming from our aim of extracting a global parameter from the fitted polarisation function, rather than targeting any individual data point. These results may find use beyond this work, and suggest that analysing noisy experimental data may be a good target for future quantum computers.\\

Nevertheless, our resource estimates highlight the challenges faced by both the algorithm introduced in this work, and many other quantum algorithms. In the context of NISQ simulations, we observed that the gate counts produced by our algorithm are likely too large to simulate on devices with realistic noise rates, even if error mitigation techniques are utilised. Similar challenges have been noted previously for NISQ electronic structure calculations~\cite{cai2019hubbard}. While noise can ultimately be overcome using quantum error correction, challenges also remain in this arena. At first glance, our algorithm appeared to require fewer fault tolerant resources than solving challenging instances of the electronic structure problem. However, the large number of repetitions required by our algorithm (to estimate observables, calculate multiple datapoints, and repeat the calculation as part of an optimisation loop) result in an impractically long runtime. While this runtime can be reduced by increasing the size of the quantum computer, or parallelising the calculation across multiple quantum machines, these both come at significant cost. We argue that this is not necessarily a limitation of our algorithm, but a challenge facing many quantum algorithms. For example, while Ref.~\cite{childs2017toward} performed a more rigorous analysis of the gate counts required to perform classically intractable simulations of spin$-\frac{1}{2}$ systems than this work, it stopped short of estimating the resources required to estimate a given observable. Their gate counts, on the order of $10^8$ T gates, are similar to those found in this work. As a result, they will run into the same problems that we have, when taking into account the resources required to estimate a given observable, or to sample a range of conditions. Similarly, while existing estimates for solving the electronic structure problem on small fault tolerant quantum proccesors~\cite{babbush2018encoding, vonburg2020MicrosoftQubitizedCatalysts,berry2019qubitizationlowrank,kivlichan2019condensedtrotter,lee2020tensor_hypercontraction_qubitization} show that we can obtain the energy of a classically intractable system in hours or days, these works typically stop short of considering what problem that actually solves. In order to optimise a molecular geometry, estimate other observables on the ground state, or elucidate a phase diagram, these calculations will likely have to be repeated many times - resulting in what may be a prohibitively long calculation time. As such, we stress that it is important to consider the total resources required to solve a given problem with a quantum computer, and not just the quantum resources required to run the corresponding circuits. The embarrassingly parallel nature of many classical algorithms (including the classical emulations in this work) and the low cost and ubiquity of classical hardware, will place stringent requirements on the performance of future quantum algorithms. Similar arguments were recently made by \textcite{babbush2020beyond_quadratic}, in the context of whether quadratic quantum speedups will be sufficient to show quantum advantage on realistic problem instances.

There are a number of optimisations that could be introduced to our algorithm, which would help to make it more practical for muon data analysis. In particular, better compilation of our time evolution circuits, or the use of more efficient time evolution algorithms, could dramatically reduce the number of gates required to achieve a given accuracy. It will also be interesting to consider which real-world systems would be most interesting to apply our algorithm to. As discussed in Sec.~\ref{Sec:MuonPolFunctions}, many systems can be satisfactorily analysed using mean-field type methods. As such, it is important to engage with the muon community to elucidate which systems appear difficult to analyse with classical techniques. One possible avenue for exploration, would be the application of our algorithm to the quantum model of muon diffusion discussed in Sec.~\ref{Sec:MuonPolFunctions} (see Eq.~(\ref{Eq:Celio_diffusion})), which has not yet been employed for data analysis, due to the large Hilbert space of the resulting simulation. Liaising with the muon community will also enable the introduction of problem-specific optimisations of the algorithm. For example, one could consider incorporating the ideas of Ref.~\cite{wilkinson2020muonfluorinefull}, which scaled the interaction strengths of more distant nuclei to act as a proxy for the rest of the sample.

A related question is whether the quantum algorithm introduced in this work can be simulated efficiently using a classical computer. The environment being in a mixed initial state may motivate the belief that approximate classical methods may be able to efficiently simulate this problem. While answering this question is beyond the scope of this work, our approach has clear links to the `one clean qubit' (DQC1) model of quantum computing~\cite{knill19981cleanqubit}, which has so far resisted efficient classical simulation~\cite{fuji2018dqc1,datta2007dqc1,morimae2014dqc1,morimae2017dqc1}.

Finally, it is interesting to ask if other, more complex quantum algorithms could be applied to analysing muon spectroscopy data. One possibility could be to use quantum read-only memory (QROM)~\cite{babbush2018encoding} to load the experimental data values into our quantum computer. We could then attempt to compute all of the simulated data in superposition (the state would resemble $\sum_t \ket{t} \ket{P(t)}$, where $t$ denotes the simulated time), and use existing quantum machine learning algorithms to extract fitting parameters of interest. While the details of this approach would likely be more complicated than outlined here, we note that this is a slightly different approach than is typically considered in quantum machine learning algorithms. Here, we exploit the exponential speedup in calculating $P(t)$ (and potentially polynomial speedups in parameter fitting), and do not mind that there is no speedup for loading in the data, as we would seek to load in a number of datapoints that is generally constant for a given type of muon spectroscopy experiment (as it is defined by the characteristics of the beam type used). Given the numerous possible avenues for exploration, we believe that both muon spectroscopy, and the analysis of data arising from other experiments underpinned by quantum mechanics, are promising targets for future quantum computers. \\ \\

\textbf{Acknowledgements\\}
This work was supported by the EPSRC National Quantum Technology
Hub in Networked Quantum Information Technology
(EP/M013243/1) and the QCS Hub (EP/T001062/1). We thank S.~Blundell and J.~Wilkinson for kindly providing the $\mu^+$SR experimental data analysed in this work, and for helpful discussions on this project, manuscript, and $\mu^+$SR experiments. We are grateful to S.~Benjamin and T.~Jones for useful discussions on this work, and for the implementation of functions for Trotterized time evolution in QuEST. We thank Z.~Cai for useful feedback and suggestions on this manuscript. We also thank X.~Yuan, Y.~Su, D.~Litinski, J.~McClean, B.~Koczor, and O.~Higgott for helpful conversations on various aspects of this work. We acknowledge the use of the University of Oxford Advanced Research Computing (ARC) facility in carrying out this work http://dx.doi.org/10.5281/zenodo.22558 .
\\

\bibliography{ChemReviewBib}

\onecolumngrid
\appendix

\section{Simulation details and additional numerical results}\label{Appendix:AdditionalResults}

The simulations presented in this work were performed using Cirq~\cite{google2020cirq} and QuEST~\cite{Jones2019quest}, as described in the main text. QuEST simulations were run on CPU nodes containing two Intel Xeon 8268 processors, each with 24 cores, with a total of 384GB of RAM. For simulations with over 21 qubits, OpenMP was used to parallelise over the 48 available cores. Each datapoint was run on a separate node, to parallelise the calculation of the polarisation function. The largest calculations, utilising 29 qubits, 30 second-order Trotter steps, and a single sample, took approximately four days to run.\\

\begin{figure*}
\includegraphics[width=1.1\columnwidth]{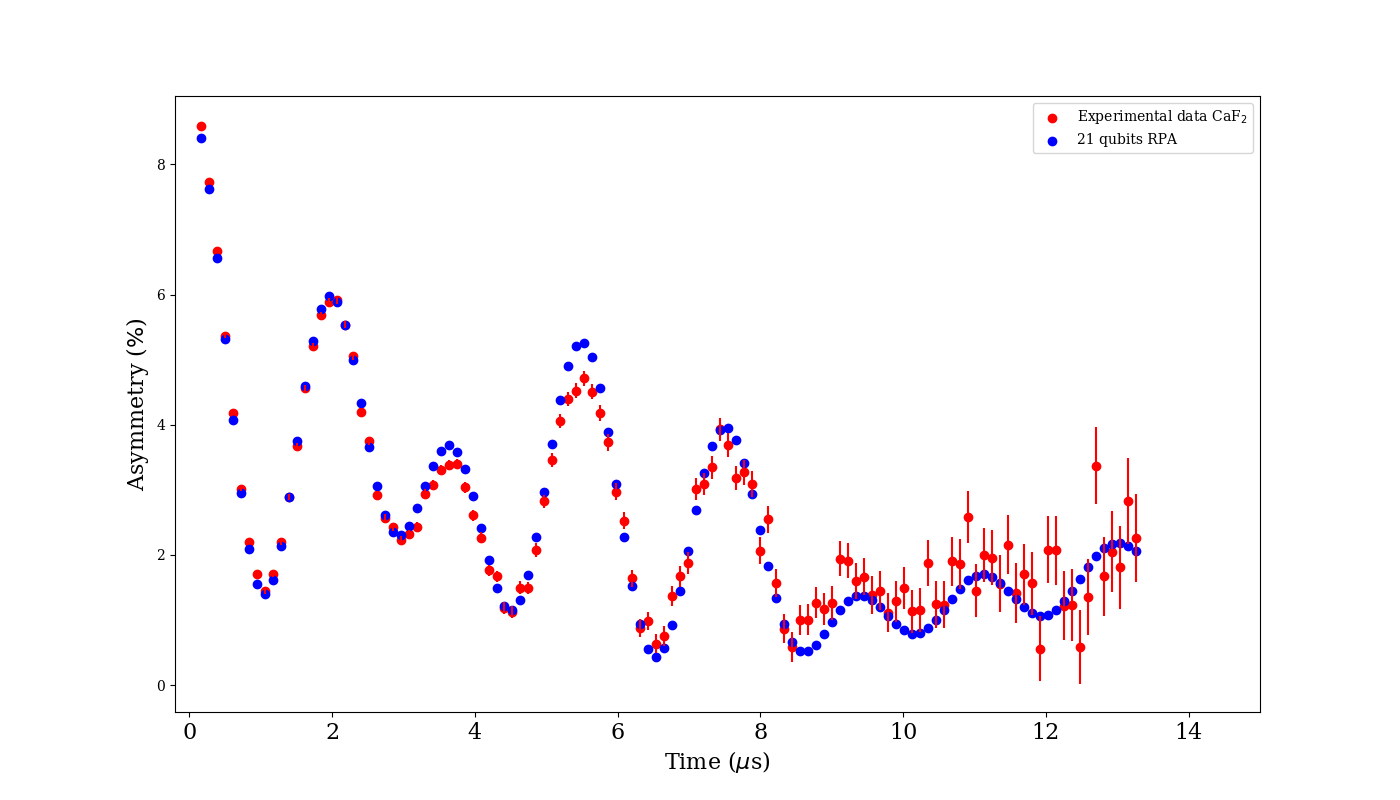}
 \caption{The CaF$_2 + \mu^+$ geometry was fixed in this simulation, and given by the density functional theory results of Ref.~\cite{wilkinson2020muonfluorinefull}. The nearest-neighbour fluorines were moved towards the muon by 0.19~\AA, and the next-nearest-neighbour fluorines were moved towards the muon by 0.027~\AA. The nn-nn fluorines were at their equilibrium positions. We used a 21 qubit simulation with the random-phase-approximation method (with 48 samples per data point, and 40 second-order Trotter steps) to generate a polarisation function for this geometry, $P(t)$. This polarisation function was fitted to the experimental asymmetry data obtained in Ref.~\cite{wilkinson2020muonfluorinefull}, using $A(t) = A_0P(t) + A_\mathrm{bg}$, where $A_0$ and $A_\mathrm{bg}$ were fitting parameters. The fit was performed using the Levenberg-Marquardt algorithm to minimise the normalised square residuals. The reduced $\chi^2$ value of the generated data is 4.44.} \label{Fig:fitting_dft}
\end{figure*}

In Fig.~\ref{Fig:fitting_dft} we present the alternative fit to the experimental data with the 21 qubit system, discussed in the main text.

\section{Preparing Dicke states on quantum computers}\label{Appendix:Dicke_States}

In this section, we discuss the method developed by \textcite{bartschi2019dicke} to prepare Dicke states on quantum computers. As discussed in the main text, to prepare a Dicke state with Hamming weight $h$, acting on $n$ qubits, the algorithm requires $\mathcal{O}(kn)$ gates, $\mathcal{O}(n)$ depth, and $n$ qubits. Here, we present a slightly less rigorous, but more pedagogical overview of the algorithm introduced in Ref.~\cite{bartschi2019dicke}, and refer the reader to the original reference for more information.

The state preparation algorithm proceeds recursively, making use of the following expression for Dicke states
\begin{equation}\label{AppEq:DickeStateRelation}
    \begin{aligned}
    \ket{D^n_h} = \sqrt{\frac{h}{n}}\ket{D^{n-1}_{h-1}}\otimes \ket{1} + \sqrt{\frac{n-h}{n}}\ket{D^{n-1}_h}\otimes \ket{0}.
    \end{aligned}
\end{equation}
We assume the existence of a unitary operator $U_{n,k}$ such that $U_{n,k} \ket{0}^{\otimes n-h} \ket{1}^{\otimes h} = \ket{D^n_h}$ for all $h \leq k$. As we show below, this operator exists, and can be constructed from typical single- and two-qubit gates.

As a first step, we note that
\begin{equation}\label{AppEq:DickeCircuitEq1}
    \begin{aligned}
    \ket{D^n_h} = U_{n,k} \ket{0}^{\otimes n-h} \ket{1}^{\otimes h},
    \end{aligned}
\end{equation}
and
\begin{equation}\label{AppEq:DickeCircuitEq2}
    \begin{aligned}
    \ket{D^n_h} =& \sqrt{\frac{h}{n}}\ket{D^{n-1}_{h-1}}\otimes \ket{1} + \sqrt{\frac{n-h}{n}}\ket{D^{n-1}_h}\otimes \ket{0}. \\
    =& \bigg{(} U_{n-1, k} \otimes I \bigg{)} \bigg{[} \sqrt{\frac{h}{n}} \ket{0}^{\otimes n-h} \ket{1}^{\otimes h} + \sqrt{\frac{n-h}{n}} \ket{0}^{\otimes n-1-h} \ket{1}^{\otimes h} \ket{0}  \bigg{]} \\
    =& \bigg{(} U_{n-1, k} \otimes I \bigg{)} \cdot \bigg{(} I_{n-k-1} \otimes V_{n,k} \bigg{)} \ket{0}^{\otimes n-h} \ket{1}^{\otimes h},
    \end{aligned}
\end{equation}
with 
\begin{equation}\label{AppEq:DickeCircuitEq2b}
    \begin{aligned}
    \bigg{(} I_{n-k-1} \otimes V_{n,k} \bigg{)} \ket{0}^{\otimes n-h} \ket{1}^{\otimes h} = \sqrt{\frac{h}{n}} \ket{0}^{\otimes n-h} \ket{1}^{\otimes h} + \sqrt{\frac{n-h}{n}} \ket{0}^{\otimes n-1-h} \ket{1}^{\otimes h} \ket{0}.
    \end{aligned}
\end{equation}
These relations imply that
\begin{equation}\label{AppEq:DickeCircuitEq3}
    U_{n,k} = \bigg{(} U_{n-1, k} \otimes I \bigg{)} \cdot \bigg{(} I_{n-k-1} \otimes V_{n,k} \bigg{)}.
\end{equation}
Here, $U_{n-1, k}$ acts on the leftmost $n-1$ qubits. $V_{n,k}$ is defined on $n$ qubits, but acts trivially on the first $n-k-1$ qubits, so can be considered to only act on the final $k+1$ qubits. We can then recurse this relationship 
\begin{equation}\label{AppEq:DickeCircuitEq4}
    U_{n-1,k} = \bigg{(} U_{n-2, k} \otimes I \bigg{)} \cdot \bigg{(} I_{n-k-2} \otimes V_{n-1,k} \bigg{)},
\end{equation}
where $V_{n-1,k}$ acts on the leftmost $k+1$ of the final $k+2$ qubits in the state. This enables us to write that 
\begin{equation}\label{AppEq:DickeCircuitEq5}
    U_{n,k} = \bigg{(} U_{k+1, k} \otimes I_{n-k-1} \bigg{)} ... \bigg{(} I_{n-k-2} \otimes V_{n-1,k} \otimes I \bigg{)} \cdot \bigg{(} I_{n-k-1} \otimes V_{n,k} \bigg{)}.
\end{equation}
We can repeat our analysis above for $U_{k+1, k}$, considering just the $k+1$ qubits acted upon:
\begin{equation}\label{AppEq:DickeCircuitEq6}
\begin{aligned}
    U_{k+1, k} \ket{0} \ket{1}^{\otimes k} &= \ket{D_k^{k+1}}\\
    &= \sqrt{\frac{k}{k+1}}\ket{D^{k}_{k-1}}\otimes \ket{1} + \sqrt{\frac{1}{k+1}}\ket{1}^{\otimes k} \otimes \ket{0} \\
    &= \bigg{(} U_{k,k-1} \otimes I \bigg{)} \bigg{[} \sqrt{\frac{k}{k+1}}\ket{0}\ket{1}^{\otimes k-1} \otimes \ket{1} + \sqrt{\frac{1}{k+1}}\ket{1}^{\otimes k} \otimes \ket{0} \bigg{]} \\ 
    &= \bigg{(} U_{k,k-1} \otimes I \bigg{)} \cdot \bigg{(} V_{k+1,k} \bigg{)} \ket{0}\ket{1}^{\otimes k-1} \otimes \ket{1},
\end{aligned}
\end{equation}
where $V_{k+1,k}$ acts on all of the $k+1$ qubits. We can then continue the recursion above
\begin{equation}\label{AppEq:DickeCircuitEq7}
\begin{aligned}
    U_{n,k} &= \bigg{(} U_{k+1, k} \otimes I_{n-k-1} \bigg{)} ... \bigg{(} I_{n-k-2} \otimes V_{n-1,k} \otimes I \bigg{)} \cdot \bigg{(} I_{n-k-1} \otimes V_{n,k} \bigg{)} \\
    &= \bigg{(} U_{k, k-1} \otimes I_{n-k} \bigg{)} \cdot \bigg{(} V_{k+1, k} \otimes I_{n-k-1} \bigg{)} ... \bigg{(} I_{n-k-2} \otimes V_{n-1,k} \otimes I \bigg{)} \cdot \bigg{(} I_{n-k-1} \otimes V_{n,k} \bigg{)} \\
    &= ... \\
    &= \bigg{(} V_{2, 1} \otimes I_{n-2} \bigg{)} \cdot \bigg{(} V_{3, 2} \otimes I_{n-3} \bigg{)} \cdot ... \cdot \bigg{(} V_{k, k-1} \otimes I_{n-k} \bigg{)} \cdot \bigg{(} V_{k+1, k} \otimes I_{n-k-1} \bigg{)} \cdot \bigg{(} I \otimes V_{k+2, k} \otimes I_{n-k-2} \bigg{)}  \cdot \\
    & ...\bigg{(} I_{n-k-2} \otimes V_{n-1, k} \otimes I \bigg{)} \cdot \bigg{(} I_{n-k-1} \otimes V_{n, k} \bigg{)}.
\end{aligned}
\end{equation}
We can express this more concisely via the expression in Lemma 2 of Ref.~\cite{bartschi2019dicke}:
\begin{equation}\label{AppEq:BartEq}
    U_{n,k} = \prod_{l=2}^k \bigg{(} V_{l,l-1} \otimes I_{n-l}  \bigg{)} \cdot \prod_{l=k+1}^n  \bigg{(} I_{l-k-1} \otimes V_{l,k} \otimes I_{n-l}  \bigg{)}.
\end{equation}
This confirms the existence of the gate $U_{n,k}$, if we are able to construct the unitary $V_{n,k}$ for arbitrary $n,k$. As discussed in Ref.~\cite{bartschi2019dicke}, we can construct $V_{n,k}$ from standard single-, two- and three-qubit gates as follows.

\begin{figure}[!h]
\begin{align*}
\Qcircuit @C=0.6em @R=.7em {
\lstick{}&\qw&\ctrl{1}&\gate{R_y(\theta)}&\ctrl{1}&\qw\\
\lstick{}&\qw&\targ&\ctrl{-1}&\targ&\qw}
\end{align*}
\caption{A circuit which yields the unitary operator given by Eq.~(\ref{AppEq:Gate1}).}\label{AppFig:Gate1}
\end{figure}
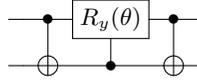

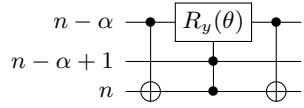
\begin{figure}[!h]
\begin{align*}
\Qcircuit @C=0.6em @R=.7em {
\lstick{n-\alpha}&\ctrl{2}&\gate{R_y(\theta)}&\ctrl{2}&\qw\\
\lstick{n-\alpha+1}&\qw&\ctrl{-1}&\qw&\qw\\
\lstick{n}&\targ&\ctrl{-2}&\targ&\qw\\}
\end{align*}
\caption{A circuit which yields the unitary operator which we refer to as $W_n^\alpha$. This operator acts trivially on all input states, except those noted in Eq.~(\ref{AppEq:Gate2}).}\label{AppFig:Gate2}
\end{figure}

We have that 
\begin{equation}\label{AppEq:MakeV}
    \begin{aligned}
    \bigg{(} V_{n,k} \bigg{)} \ket{0}^{\otimes k+1-h} \ket{1}^{\otimes h} = \sqrt{\frac{h}{n}} \ket{0}^{\otimes k+1-h} \ket{1}^{\otimes h} + \sqrt{\frac{n-h}{n}} \ket{0}^{\otimes k-h} \ket{1}^{\otimes h} \ket{0}.
    \end{aligned}
\end{equation}
This unitary only changes the value of the zeroth qubit from the right, and the $h$th qubit from the right. First, consider the circuit shown in Fig.~\ref{AppFig:Gate1}. If we set $\theta = 2\mathrm{cos^{-1}}(\sqrt{1/n})$, then the circuit applies the unitary
\begin{equation}\label{AppEq:Gate1}
\begin{bmatrix}
1& 0 & 0 & 0\\
0& \sqrt{\frac{1}{n}} & -\sqrt{\frac{n-1}{n}} & 0\\
0& \sqrt{\frac{n-1}{n}} & \sqrt{\frac{1}{n}} & 0\\
0& 0 & 0 & 1
\end{bmatrix}.
\end{equation}
Similarly, the circuit in Fig.~\ref{AppFig:Gate2} with $\theta = 2\mathrm{cos^{-1}}(\sqrt{\alpha/n})$ acts trivially on all input states, except for
\begin{equation}\label{AppEq:Gate2}
    \begin{aligned}
    \ket{011} \rightarrow \sqrt{\frac{\alpha}{n}} \ket{011} + \sqrt{\frac{n - \alpha}{n}} \ket{110},\\
    \ket{110} \rightarrow -\sqrt{\frac{n-\alpha}{n}} \ket{011} + \sqrt{\frac{\alpha}{n}} \ket{110}.    
    \end{aligned}
\end{equation}
We can now use these building blocks to construct the unitary $V_{n,k}$. We first apply the gate in Fig.~\ref{AppFig:Gate1} to the final two qubits in the register, and then repeatedly apply the gate $W_n^\alpha$ shown in Fig.~\ref{AppFig:Gate2}, incrementing the value of $\alpha$ each time. As we are applying these gates on the state $\ket{0}^{\otimes n-h} \ket{1}^{\otimes h}$, the first $h$ qubits encountered are all in the $\ket{1}$ state, so these gates act trivially. After these $h$ qubits, the three qubit $W_n^h$ gate encounters the state $\ket{0}_{n-h}\ket{1}_{n-h+1}\ket{1}_n$, which it transforms into $\sqrt{\frac{h}{n}} \ket{0}_{n-h}\ket{1}_{n-h+1}\ket{1}_n + \sqrt{\frac{n-h}{n}} \ket{1}_{n-h}\ket{1}_{n-h+1}\ket{0}_n$. The subsequent $W_n^{i>h}$ gates act trivially on both branches of this superposition. Taken together, these gates carry out the transform in Eq.~(\ref{AppEq:MakeV}), for all $h \leq k$. It is shown in Ref.~\cite{bartschi2019dicke} how the three-qubit $W$ gate can be decomposed into single- and two-qubit gates, and how the circuit can be implemented with $\mathcal{O}(kn)$ gates, $\mathcal{O}(n)$ depth, and $n$ qubits on a linear array of qubits, with nearest-neighbour connectivity.

\end{document}